\documentclass{jfm}
\usepackage{graphicx}
\usepackage{epstopdf, epsfig}

\usepackage{subcaption}
\newcommand{\hi}{\mathrm{i}}
\newcommand{\hr}{\mathrm{r}}

\usepackage{cleveref}

\crefformat{section}{\S#2#1#3}

\shorttitle{Effects of Distributed Roughness on Crossflow Instability}
\shortauthor{J. He, A. Butler and X. Wu}

\title{Effects of Distributed Roughness on Crossflow Instability through Generalized Resonance Mechanisms}

\author{Jiyang He\aff{1},
        Adam Butler\aff{2}
 \and  Xuesong Wu\aff{1}\aff{,}\aff{2}
  \corresp{\email{x.wu@ic.ac.uk}}}

\affiliation{\aff{1}Department of Mechanics, Tianjin University, Tianjin 300072, China
\aff{2}Department of Mathematics, Imperial College London, 180 Queen’s Gate, London SW7 2AZ, UK}

\begin{document}

\maketitle

\begin{abstract}
Experiments have shown that micron-sized distributed surface roughness can significantly promote transition in a three-dimensional boundary layer dominated by crossflow instability. This sensitive effect has not yet been fully explained physically and mathematically. Unlike past researches focusing on the receptivity of the boundary layer to surface roughness, or on the local stability of the modified mean flow, this paper seeks possible inherent mechanisms by investigating the effects of distributed surface roughness on crossflow instability through resonant interactions with eigenmodes. A key observation is that the perturbation induced by roughness with specific wavenumbers can interact with two eigenmodes (travelling and stationary vortices) through triadic resonance, or interact with one eigenmode (stationary vortices) through Bragg scattering. Unlike the usual triadic resonance of neutral, or nearly neutral, eigenmodes, the present triadic resonance can take place among modes with \textit{O}(1) growth rates, provided that these are equal; unlike the usual Bragg scattering involving neutral waves, crossflow stationary vortices can also be unstable. For these amplifying waves, the generalized triadic resonance and Bragg scattering are put forward, and the resulting corrections to the growth rates are derived by a multiple-scale method. The analysis is extended to the case where up to four crossflow vortices interact with each other in the presence of suitable roughness components. The numerical results for Falkner-Skan-Cooke boundary layers show that roughness with a small height (a few percent of the local boundary-layer thickness) can change growth rates substantially (by a more-or-less \textit{O}(1) amount). This sensitive effect is attributed to two facts: $(a)$ the resonant nature of the triadic interaction and Bragg scattering, which makes the correction to the growth rate proportional to the roughness height, and $(b)$ the wavenumbers of the roughness component required for the resonance are close to those of the neutral stationary crossflow modes, as a result of which a small roughness can generate a large response.
Another important effect of roughness is that its presence renders the participating eigenmodes, which are otherwise independent, fully coupled. Our theoretical results suggest that micron-sized distributed surface roughness influences significantly both the amplification and spectral composition of crossflow vortices.
\end{abstract}


\section{Introduction}
It is known that laminar-turbulence transition in boundary layers is significantly influenced by surface roughness. The latter may be localised or distributed in one or both of the streamwise and spanwise directions. Depending on their location, length scale and height, surface roughness elements may influence transition through different mechanisms, which include, as we shall discuss in more detail, receptivity, alteration of the base flow thereby changing the instability characteristics and local scattering mechanism. 
Before proceeding to theoretical studies of the mechanisms, experimental observations about roughness are reviewed first.

For two-dimensional (2D) boundary layers, which are, in the absence of roughness, susceptible to the viscous Tollmien-Schlichting (T-S) instability, \citet{klebanoff1972mechanism} observed that earlier transition was caused by an isolated 2D roughness (a cylindrical rod) with $h^*/\delta^*=0.7-0.8$, where $h^*$ is the roughness height and $\delta^*$ the boundary-layer displacement thickness. They argued convincingly  that the flow is strongly destabilized in the separated region downstream of the roughness, where the inflectional profile is susceptible to a strong inviscid instability. \citet{corke1986experiments} studied the instability of 2D boundary layers with and without distributed roughness (sand paper) of the height $h^*/\delta^*\approx0.5$. Although greatly enhanced growth of T-S waves was observed, the boundary-layer profile over a rough wall did not become inflectional, and kept a similar shape factor as that on the smooth wall. Therefore, the increased growth of T-S fluctuations was not attributed to inviscid instability, but to the continual excitation of T-S waves on the rough wall by free-stream turbulence.

 \citet{ma2014effects} investigated the effects of sinusoidal surface corrugations with wavelengths of the same order-of-magnitude as that of the T-S waves in a zero-pressure-gradient boundary layer. They observed significantly enhanced growth of T-S waves by 2D corrugation even though its amplitude was only about $10\%$ of the displacement thickness. In contrast, three-dimensional (3D) corrugations had only a small influence even for a larger amplitude of $20\%$ of the displacement thickness. On the other hand, 3D corrugations could promote a secondary instability causing the development of the Klebanoff-type peak-valley aligned structure. \citet{hamed2016transition} carried out a laboratory investigation of transition over 2D and 3D periodic large-scale roughnesses. The result showed that the flow over the 2D roughness becomes turbulent much sooner than its 3D counterpart. This is likely due to the fact that flow separation (which leads to an inflectional profile) is less severe in the 3D case. 

In 3D boundary layers, crossflow instability operates, leading to amplification of stationary and travelling vortices \citep{saric2003stability}. A series of wind-tunnel experiments on the effects of roughness on transition have been conducted since 1990s. \citet{reibert1996experiments} and \citet{carrillo1997distributed} noted that naturally present surface irregularities with height in microns (or submicron when polished) trigger stationary vortices, which grow to cause transition.
\citet{radeztsky1999effect} studied the impact of micron-sized distributed and isolated roughness elements on transition in the boundary layer over a swept wing. For the case of the chordwise Reynolds number $2.7\times10^6$, as the surface was polished and roughness height was reduced from $9.0\mu m$ to $0.25\mu m$, the transition location was delayed from $40\%$ to $68\%$ of the chord length. This phenomenon indicates extreme sensitivity to micron-sized distributed roughness. On the other hand, when isolated  roughness elements with $6\mu m$ height were placed near the attachment line, earlier transition was observed, which was attributed to stronger stationary vortices being excited by the roughness through receptivity. \citet{radeztsky1999effect} further pointed out that ``distributed roughness (random) appears to have the dangerous scales that effect transition". \citet{carrillo1997distributed}  demonstrated experimentally that submicron-high spanwise-periodic distributed roughness elements (DREs) applied near the attachment line can influence transition significantly. In particular, DREs with a spanwise spacing shorter than the wavelength of the linearly most unstable vortices delay transition. This surprising effect together with the high sensitivity to roughness led \citet{saric1998leading} to suggest that DREs could be an effective laminar-flow-control technique. Its effectiveness has been assessed in a series of wind-tunnel and flight-test experiments \citep{carpenter2008laminar,carpenter2009flight,woodruff2011receptivity}. \citet{woodruff2011receptivity} noted that the effectiveness of DREs is dependent on the free-stream turbulence level.

The theoretical frameworks that account for the effects of surface roughness on boundary-layer transition can, according to the physical mechanisms being described, be divided into three categories: receptivity, instability and local scattering. Receptivity refers to the process in which external disturbances generate instability modes in the boundary layer \citep{goldstein1989boundary,saric2002boundary}. In order for roughness elements to excite instability modes, they must have a lengthscale comparable with the characteristic wavelength of the latter. If the modes to be excited are time-dependent (e.g. T-S waves in 2D boundary layers and travelling vortices in 3D boundary layers), the roughness-induced steady perturbation needs to interact with free-stream unsteady perturbations within a suitable frequency band, in order to produce the required forcing. In contrast, roughness in 3D boundary layers with suitable lengthscale can itself induce stationary crossflow vortices. 
\citet{crouch1993receptivity} and \citet{choudhari1994roughness} studied the receptivity of crossflow stationary modes to surface roughnesses using a finite-Reynolds-number theory (FRNT), which consists of the Orr-Sommerfeld (O-S) equation with the wall roughnesses being accounted for through inhomogeneous conditions. Using FRNT, \citet{ng1999roughness} calculated the initial amplitude pertaining to the experiment of \citet{reibert1996experiments}, and the comparison with the measurements suggests that this FRNT works quite well. All of these studies neglected non-parallelism, which \citet{bertolotti2000receptivity} set to include by expanding the base flow as a Taylor series. The FRNT based on the O-S equation is applicable only to roughness of sufficiently small height. Receptivity to large-height (nonlinear) roughness was considered by \citet{choudhari1996nonlinear} using the nonlinear triple-deck theory. Receptivity to surface roughness has also been studied by direct numerical simulations (DNS). \citet{collis1999receptivity} compared the predictions by DNS and FRNT, and found that neglecting non-parallel-flow effects led to over-prediction of the initial amplitudes, a conclusion that was later confirmed by \citet{Schrader2009receptivity}. \citet{tempelmann2012swept} performed DNS of the receptivity process in the experiment of \citet{reibert1996experiments}. The predicted amplitude is about $40\%$  of the measurement, and the discrepancy was attributed to an experimental uncertainty of the height and streamwise location of the roughness elements (micro-cylinders). \citet{hosseini2013stabilization} investigated the receptivity to the DREs introduced near the leading edge for transition control, and their simulations confirmed the results of earlier experiments \citep{saric1998leading}. 
The receptivity pertaining to the flight-test experiments \citep{carpenter2009flight} was modeled by \citet{rizzetta2010direct} by DNS, however, no comparison was made with experimental data. \citet{kurz2014receptivity} simulated receptivity to micron-sized roughness elements, which were meshed, and found that the receptivity coefficient (the ratio of the amplitude of the stationary vortices excited to the roughness height $h$) is not a constant as was often asserted, but a linear function of $h$. This behaviour, referred to as ``superlinearity", was shown theoretically to be an important feature of the receptivity near the leading edge \citep{butler2018stationary}.

Surface roughness can also influence transition by altering instability characteristics. Theoretical studies about this involve computing the roughness-modified mean flow and performing linear stability analysis of this new state. The effect of an isolated roughness (hump) in a 2D boundary layer was considered by \citet{nayfeh1988effect}, \citet{cebeci1989prediction}, \citet{masad1994transition} and \citet{gao2011stability}. In order to capture possible flow separation created by  the hump, the roughness-modified mean flow was computed by solving the interactive boundary-layer (IBL) equations or by DNS. Its linear stability was then studied based on the O-S equation or PSE (parabolic stability equations) \citep{gao2011stability,park2013linear}. Overall, the growth of T-S waves was enhanced. Clearly, for the PSE or the O-S equation to be applicable the streamwise lengthscale of the hump must be much longer than the characteristic T-S wavelength. 

Similar studies have been carried out for distributed roughness (wavy walls). \citet{lessen1976effect} first calculated the response of a 2D boundary layer to small-amplitude surface waviness by solving the steady O-S equation subject to inhomogeneous boundary conditions. The resulting signature was used to evaluate the Reynolds stresses that modify the mean flow. The latter was obtained by solving the Reynolds-averaged boundary-layer equations. A linear stability analysis, based on the O-S equation for the modified mean flow, was performed. The effect is of \textit{O}($h^2$), nevertheless, the calculation showed a $10\%$ reduction of the minimum critical Reynolds number in the presence of roughness with a height of only $1\%$ of the boundary-layer thickness. \citet{gaster2016understanding} also computed the mean flow over a wavy wall using the O-S formulation. A linear stability analysis indicated increased amplification rates of T-S waves. Effects of wavy walls with a large enough amplitude to cause pockets of local separation were investigated by \citet{wie1998effect} and \citet{thomas2017development}, who used respectively an IBL approach and a Navier-Stokes solver to compute the roughness-modified mean flow. The development of T-S waves was then studied by the PSE approach. A wavy wall was found to play a destabilising role in general. \citet{thomas2017development} found that shorter waviness has a stronger destabilising effect, but some long-wavelength deformations create a favorable pressure gradient to damp the growth of T-S waves.

For 3D boundary layers, the effect of localised surface waviness on transition was studied by \citet{masad1996effect}. The IBL approach was used to calculate the distorted mean flow, for which a linear stability analysis was performed. \citet{thomas2016stability} considered extended chordwise surface waviness. The mean flow was computed by a Navier-Stokes solver, and the instability of the distorted flow was assessed by PSE and linearized Navier-Stokes methods. Both studies found that wavy walls enhanced the amplification of crossflow vortices. Note that the roughness heights considered by \citet{thomas2016stability} were larger than $10\%$ of the displacement thickness, and their study was not pertaining to the experimental phenomenon for micron-sized distributed roughness.  

\citet{floryan1997stability} investigated the effects of (simulated) wavy walls with short wavelengths comparable with the shear-layer thickness. Considering the spatial periodicity of the mean flow, Floquet theory was applied to study the stability. A new type of instability characterized by streamwise vortices as the dominant component was found. The usual T-S waves were found to be moderately destabilised. \citet{cabal2002stability} used a similar methodology to study the stability of the flow in a wavy channel, and found two types of instability modes: streamwise vortices induced by the wall surface and T-S waves modified by the presence of wall waviness. The former might be caused by centrifugal effects over a wavy wall \citep{floryan2002centrifugal}.
  
A different scenario occurs when a localized roughness is located in the main unstable region. An incoming instability wave will be scattered as it propagates through the streamwise inhomogeneous region created by the roughness. Of special interest is the roughness with lengthscale comparable with the characteristic wavelength of the instability, in which case neither O-S equation nor PSE is valid due to the strong non-parallelism. Mathematically, this new mechanism of scattering is elliptic. A local scattering theory based on triple-deck framework was proposed by \citet{wu2016local}.  The impact of the localised roughness on transition was characterized naturally by a transmission coefficient, defined as the ratio of the amplitude of the T-S wave downstream of the roughness to that upstream. Their results showed that roughness played a destabilising role. A finite-Reynolds-number formulation of the local scattering theory was developed by \citet{huang2017local}, and was verified by DNS results and experimental data, with which good agreement was obtained.

In the present study, we seek possible mechanisms which underpin the extreme sensitivity of crossflow instability and transition to micron-sized distributed roughness.  Unlike existing explanations from the perspective of receptivity or local instability, we try to understand the phenomenon from the standpoint of resonant interactions between instability modes and the roughness-induced perturbation. Three-dimensional distributed roughness is modeled in the simplest case by a single Fourier component, which is periodic in both the chordwise and spanwise directions, or more realistically by a superposition of such components. The perturbation induced by each roughness component will be referred to as `roughness mode' herein despite the fact that it is not an eigenmode. We identify the components that can participate in some forms of resonant interactions with suitable crossflow eigenmodes, thereby changing the stability most effectively. Unlike previous stability analyses for the roughness-modified mean flow, which ignored wave-wave interaction, we retain the base flow of a smooth wall, but let the disturbance be composed of roughness modes and crossflow eigenmodes. The resonant interactions can occur if their wavenumbers and frequencies satisfy certain resonance conditions, and these include triadic resonance between two crossflow eigenmodes and one roughness mode, and Bragg scattering involving one stationary crossflow eigenmode and one roughness mode. Triadic resonance is a general mechanism in wave mechanics, but was first proposed by \citet{craik1971non} to describe the nonlinear instability of T-S waves. It is usually required that the interacting waves have small growth rates, i.e. they are (nearly) neutral. The resonant triad in our study is a generalized one in that the crossflow eigenmodes involved can have \textit{O}(1) growth rates, and all that is needed is that their growth rates are (nearly) the same. Bragg scattering, which originated from crystallography, is a general wave phenomenon or mechanism as well. It  was invoked, e.g., to explain the strong reflection of water waves induced by periodic sandbar \citep{mei1985resonant,mei1988note}, where the wavenumber and frequency of water waves are always real. Bragg scattering is generalized in our work to study the interaction of a roughness mode with an unstable crossflow eigenmode. Unlike local stability analysis, which requires the roughness wavelength to be much larger than that of instability modes, the wavelengths of the roughness modes and crossflow eigenmodes in our resonant interactions can be comparable. Furthermore, the resonant interactions considered here involve multiple eigenmodes, whereas linear stability analysis involves only one eigenmode.  

The rest of the paper is organized as follows. The problem is formulated in \cref{sec:formulation}, where after specifying the base flow, we present the homogeneous and inhomogeneous boundary-value problems governing the crossflow eigenmodes and roughness modes respectively, and in particular the generalized resonance mechanisms are explained with reference to the dispersion relations of the crossflow vortices. The mathematical description of the resonant interactions is presented in \cref{sec:derivation}. Using a multi-scale method, we derive the amplitude equations governing the interacting vortices; these equations allow us to compute the corrections to the growth rates. Numerical results are shown in \cref{sec:results} for stationary and travelling crossflow vortices. Finally, the conclusions are summarized and topics of future investigations are discussed in \cref{sec:conclusions}.

\section{Problem formulation}\label{sec:formulation}
\subsection{Governing equations}
We consider the three-dimensional incompressible boundary layer over a surface (of a swept wing or wedge), on which small-amplitude distributed surface roughness is present. The flow is described in the Cartesian coordinate system ($x^*$, $y^*$, $z^*$) with its origin at the leading edge, where $x^*$, $y^*$ and $z^*$ are in the chordwise, vertical and spanwise directions respectively. The coordinates and velocities are non-dimensionalized by the local boundary-layer displacement thickness $\delta^*_0$ and chordwise slip velocity $U^*_{\infty,0}$ at the chordwise location $x^*_0$ where the resonant interaction occurs, and $x^*_0$ will be also referred to as the resonant point. The Reynolds number is defined as
\begin{equation}
Re=U^*_{\infty,0}\delta^*_0/\nu^*,
\end{equation}
where $\nu^*$ is the kinematic viscosity.

The flow is governed by the non-dimensional incompressible Navier-Stokes (N-S) equations and continuity equation,
\begin{equation}\label{eq:NS}
\frac{\partial \boldsymbol{u}}{\partial t}+(\boldsymbol{u}\cdot\nabla)\boldsymbol{u}+\nabla p-\frac{1}{Re}\nabla^2\boldsymbol{u}=0, \quad \nabla\cdot\boldsymbol{u}=0,
\end{equation}
where $\boldsymbol{u}=(u, v, w)$ with $u$, $v$ and $w$ standing for the chordwise, wall-normal and spanwise velocities respectively, and $p$ denotes the pressure non-dimensionalized by $\rho^*U^{*2}_{\infty,0}$ with $\rho^*$ being the reference density. The velocity and pressure, $(\boldsymbol{u}, p)$, are decomposed into a base flow $(\boldsymbol{U}, P)$ and a disturbance part $(\boldsymbol{u'}, p')$,
\begin{equation}\label{eq:decomposition}
\boldsymbol{u}=\boldsymbol{U}+\boldsymbol{u'}, \quad p=P+p'.
\end{equation}
The base flow is assumed to be spanwise uniform. Its velocity field $\boldsymbol{U}=(U, V, W)$ satisfies the non-dimensional boundary-layer equations,
\begin{subeqnarray}\label{eq:boundary_layer_equation}
U\frac{\p U}{\p x}+V\frac{\p U}{\p y}&=&U_{\infty}\frac{\mathrm{d}U_{\infty}}{\mathrm{d}x}+\frac{1}{Re}\frac{\p^2 U}{\p y^{2}}, \\
U\frac{\p W}{\p x}+V\frac{\p W}{\p y}&=&\frac{1}{Re}\frac{\p^2 W}{\p y^{2}}, \\
\frac{\p U}{\p x}+\frac{\p V}{\p y}&=&0,
\end{subeqnarray}
and the boundary conditions,
\begin{equation}\label{eq:boundary_layer_bc}
U=V=W=0 \quad \mbox{at\ }\quad y=0; \quad U\to U_\infty, \quad W \to W_\infty \quad \mbox{as\ }\quad y\to \infty,
\end{equation}
where $U_\infty$ and $W_\infty$ stand for the non-dimensional slip velocities along the chordwise and spanwise directions respectively.
Substituting the decomposition (\ref{eq:decomposition}) into (\ref{eq:NS}) and subtracting out the equations governing the base flow lead to the nonlinear disturbance equations,
\begin{equation}\label{eq:nonlinear_disturbance_equations}
\frac{\partial \boldsymbol{u}}{\partial t}+(\boldsymbol{U}\cdot\nabla)\boldsymbol{u}+(\boldsymbol{u}\cdot\nabla)\boldsymbol{U}+\nabla p-\frac{1}{Re}\nabla^2\boldsymbol{u}=-(\boldsymbol{u}\cdot\nabla)\boldsymbol{u}, \quad \nabla\cdot\boldsymbol{u}=0.\\
\end{equation}
Without ambiguity, the prime symbol for disturbance quantities is omitted for simplicity.

\subsection{Base flow}
For simplicity, the base flow is taken to be the Falkner-Skan-Cooke (FSC) boundary layer, which has frequently been used as a convenient vehicle to investigate key receptivity, instability and transition mechanisms of three-dimensional boundary layers \citep[e.g.][]{hogberg1998secondary,Schrader2009receptivity}. The chordwise and spanwise slip velocities are given by
\begin{equation}
\frac{U_{\infty}^*(x^*)}{U_{\infty}^*(x^*_0)}=\left(\frac{x^*}{x^*_0}\right)^m,\quad W_{\infty}^*=\mathrm{const},
\end{equation}
where the acceleration parameter $m$ is related to the Hartree parameter $\beta_H$ through the relation $m=\beta_H/(2-\beta_H)$. The local sweep angle $\phi_0$ at the resonant point $x^*_0$ is defined as $\phi_0=\tan^{-1}(W^*_{\infty}/U^*_{\infty}(x^*_0))=\tan^{-1}(W_{\infty})$ on noting that $U_{\infty}(x_0)=1$. The values of $m$ and $\phi_0$ (or equivalently $\beta_H$ and $W_{\infty}$) can be chosen such that the profiles are representative of the local flow in the boundary layer over a swept wing.

This boundary-layer flow admits a similarity solution. Introducing a similarity variable $\eta$ and a non-dimensional stream function $\Psi$,
\begin{equation}
\eta=\sqrt{\frac{m+1}{2}\frac{U_{\infty}Re}{x}}y, \quad \Psi=\sqrt{\frac{2}{m+1}\frac{U_{\infty}x}{Re}}f(\eta),
\end{equation}
we may express the base-flow velocities as $U=\partial \Psi/\partial y$, $V=-\partial \Psi/\partial x$ and $W=W_{\infty}g(\eta)$, substitution of which into (\ref{eq:boundary_layer_equation}) and (\ref{eq:boundary_layer_bc}) leads to the ordinary differential equations
\begin{equation}
f^{\prime\prime\prime}+ff^{\prime\prime}+\beta_H(1-f^{\prime2})=0,\quad g^{\prime\prime}+fg^{\prime}=0,
\end{equation}
and the boundary conditions
\begin{equation}
f=f^{\prime}=g=0 \quad \mbox{at\ }\quad \eta=0; \quad f^{\prime}=g^{\prime}=1 \quad \mbox{at\ }\quad \eta=\infty,
\end{equation}
where the symbol $'$ stands for the derivative with respect to $\eta$.

By the standard definition, the boundary-layer displacement thickness is expressed as
\begin{equation}
\delta_0^*=\left(\frac{m+1}{2}\frac{U_{\infty}^*(x^*_0)}{\nu^*x^*_0}\right)^{-\frac{1}{2}}c, \quad c=\int_0^{\infty}(1-f^{\prime})\mathrm{d}\eta.
\end{equation}
Non-dimensionalizing the first of the above equations gives the relation between the non-dimensional resonant point $x_0$ and the corresponding Reynolds number $Re$, $x_0=(m+1)Re/(2c^2)$ \citep{hogberg1998secondary}.
% \begin{equation}\label{reference_location}
% x_0=\frac{(m+1)Re}{2c^2}.
% \end{equation}

The dimensionless velocities within the boundary layer and their derivatives (with respect to $y$) can be expressed as
\begin{equation}
U(y)=f^{\prime}(\eta),\quad \frac{\mathrm{d}U}{\mathrm{d}y}=cf^{\prime\prime}(\eta);  \quad W(y)=W_{\infty}g(\eta),\quad \frac{\mathrm{d}W}{\mathrm{d}y}=W_{\infty}cg^{\prime}(\eta). 
\end{equation}

\subsection{Crossflow eigenmodes}
According to usual linear stability theory, crossflow eigenmodes, which may be present in the absence of roughness,  take the normal form,
\begin{equation}\label{eq:eigenmode}
(u,v,w,p)=\epsilon(\hat{u},\hat{v},\hat{w},\hat{p})E+c.c., \quad E=e^{\mathrm{i}(\alpha x+\beta z-\omega t)},
\end{equation}
where $\alpha$ and $\beta$ are the chordwise and spanwise wavenumbers respectively, $\omega$ is the frequency and $\epsilon\ll1$ is a measure of the amplitude. If the Reynolds number $Re$ is taken to be asymptotically large, a variety of long-wavelength regimes (with $\alpha, \beta \ll1$) arises \citep{choudhari1995long}. In the present work, we treat $Re$ as being finite, but make the usual parallel-flow approximation, with which substitution of (\ref{eq:eigenmode}) into (\ref{eq:nonlinear_disturbance_equations}) leads to a system of the first-order equations, 
\begin{equation}\label{eq:one_order_OS}
\frac{\mathrm{d} \hat{\boldsymbol{\varphi}}}{\mathrm{d}y}=\mathsfbi{L}\hat{\boldsymbol{\varphi}},
\end{equation}
where the vector $\hat{\boldsymbol{\varphi}}=[\hat{u},\hat{v},\hat{w},D\hat{u},\hat{p},D\hat{w}]^T$, and the operator $\mathsfbi{L}$ is given by
\begin{equation}\label{eq:matrix}
\setlength{\arraycolsep}{0pt}
\renewcommand{\arraystretch}{1.3}
 \mathsfbi{L}=
 \left[
\begin{array}{cccccc}
0\quad&0\quad&0\quad&1\quad&0\quad&0\\
-\hi\alpha\quad&0\quad&-\hi\beta\quad&0\quad&0\quad&0 \\
0\quad&0\quad&0\quad&0\quad&0\quad&1\\
ReS\quad&ReDU\quad&0\quad&0\quad&\hi\alpha Re\quad&0\\
0\quad&-S\quad&0\quad&-\hi\alpha/Re\quad&0\quad&-\hi\beta/Re\\
0\quad&ReDW\quad&ReS\quad&0\quad&\hi\beta Re\quad&0
\end{array}  \right], \nonumber
\end{equation}
with $S=-\hi\omega+\hi\alpha U+\hi\beta W+(\alpha^2+\beta^2)/Re$ and $D$ standing for $\mathrm{d}/\mathrm{d}y$.
The boundary conditions follow from the requirements that the perturbation velocities
vanish at the smooth wall and decay to zero in the free stream, and so they can be expressed as
\begin{equation}\label{eq:os_boundary_conditions}
\hat{u}(y)=\hat{v}(y)=\hat{w}(y)=0 \quad \mbox{at\ }\quad y=0; \quad \hat{u}(y), \hat{v}(y), \hat{w}(y)\rightarrow 0 \quad \mbox{as\ } \quad y\rightarrow\infty.
\end{equation}
By eliminating $\hat{p}$, $\hat{u}$ and $\hat{w}$, the system (\ref{eq:one_order_OS}) can be reduced to the well-known O-S equation for $\hat{v}$,
\begin{equation}\label{eq:os}
\Big[ (-\mathrm{i}\omega+\mathrm{i}\alpha U+\mathrm{i}\beta W)(D^2-\alpha^2-\beta^2)-\mathrm{i}\alpha U^{\prime\prime}-\mathrm{i}\beta W^{\prime\prime}-\frac{1}{Re}(D^2-\alpha^2-\beta^2)^2 \Big] \hat{v}=0,
\end{equation}
 but we will solve the eigenvalue problem (\ref{eq:one_order_OS})-(\ref{eq:os_boundary_conditions}) in the primitive variables, for which an efficient algorithm exists \citep{malik1990numerical}. The present finite-Reynolds-number formulation represents a composite theory valid for all long-wavelength (and possibly viscous) regimes \citep{choudhari1995long} except the lower-branch regime, where non-parallelism is a leading-order effect \citep{butler2018stationary}.
 We are concerned with spatial stability, and so the chordwise wavenumber $\alpha$ is calculated for a given spanwise wavenumber $\beta$ and  frequency $\omega$. Usually $\alpha=\alpha_\hr+\hi\alpha_\hi$ is complex with $-\alpha_\hi$ representing the growth rate. The instability modes are referred to as travelling vortices if $\omega\neq0$ and stationary vortices if $\omega=0$.

The eigenvalue problem posed by equation (\ref{eq:one_order_OS}) and the boundary conditions (\ref{eq:os_boundary_conditions}) is solved numerically. The equation is discretized by using the fourth-order compact finite-difference scheme of \citet{malik1990numerical}. The resulting algebraic system has non-zero solutions only when the determinant of its coefficient matrix is equal to zero. For a given $\alpha$, $\beta$ and $\omega$, the determinant is computed by Gaussian elimination, and an appropriate $\alpha$ making the determinant vanish is found by M$\ddot{\mathrm{u}}$ller iteration.

\subsection{Roughness modes}
In general, the wall shape can be represented as a superposition of Fourier components,
\begin{equation}\label{eq:multiple_wavenumber_roughness}
y_w=\sum\limits_{\alpha_w,\beta_w}h(\alpha_w, \beta_w)e^{\hi(\alpha_wx+\beta_wz)}+c.c.,
\end{equation}
where $\alpha_w$ and $\beta_w$ are the chordwise and spanwise roughness wavenumbers respectively. The non-dimensional height of the distributed roughness is assumed to be small enough that a perturbation scheme is deemed applicable. Then each mode is independent, and we can calculate the boundary-layer response to each Fourier component,
\begin{equation}\label{wavy_wall}
y_w=he^{\hi(\alpha_wx+\beta_wz)}+c.c.,
\end{equation}
which is to be referred to as single-wavenumber roughness, and a more general form (\ref{eq:multiple_wavenumber_roughness})  will be referred to as a multiple-wavenumber roughness.

Under the local parallel-flow assumption, the perturbation induced by (\ref{wavy_wall}) takes the form,
\begin{equation}\label{eq:roughness_mode_form}
[u_w,v_w,w_w,p_w]=h[\hat{u}_w(y), \hat{v}_w(y), \hat{w}_w(y), \hat{p}_w(y)]e^{\mathrm{i}(\alpha_wx+\beta_wz)}+c.c.,
\end{equation}
where $\hat{u}_w(y)$ and etc. are shape functions. Introducing (\ref{eq:roughness_mode_form}) into (\ref{eq:nonlinear_disturbance_equations}) leads to the O-S equation for $\hat{v}_w$,
\begin{equation}\label{eq:statinary_os}
\Big[ (\mathrm{i}\alpha_w U+\mathrm{i}\beta_w W)(D^2-\alpha_w^2-\beta_w^2)-\mathrm{i}\alpha_w U^{\prime\prime}-\mathrm{i}\beta_w W^{\prime\prime}-\frac{1}{Re}(D^2-\alpha_w^2-\beta_w^2)^2 \Big] \hat{v}_w=0,
\end{equation}
or equivalently the first-order system,
\begin{equation}\label{eq:os_forced}
\frac{\mathrm{d} \hat{\boldsymbol{\varphi}}_w}{\mathrm{d}y}=\mathsfbi{L}_w\hat{\boldsymbol{\varphi}}_w,
\end{equation}
where $\hat{\boldsymbol{\varphi}}_w$ and $\mathsfbi{L}_w$ have similar expressions as $\hat{\boldsymbol{\varphi}}$ and $\mathsfbi{L}$ in (\ref{eq:one_order_OS})-(\ref{eq:matrix}) for crossflow eigenmodes. 
At the wavy wall, the no-slip condition must be satisfied. By Taylor expansion, shifting the boundary to $y=0$ leads to the inhomogeneous boundary conditions,
\begin{equation}\label{eq:os_forced_y0}
\hat{u}_w(0)=-U'(0),\quad\hat{v}_w(0)=0,\quad\hat{w}_w(0)=-W'(0).
\end{equation}
The boundary conditions in the far field are
\begin{equation}\label{eq:os_forced_y1}
\hat{u}_w(y), \hat{v}_w(y), \hat{w}_w(y)\rightarrow 0 \quad \mbox{as\ } \quad y\rightarrow\infty.
\end{equation}
Discretization of equation (\ref{eq:os_forced}) and its boundary conditions (\ref{eq:os_forced_y0})-(\ref{eq:os_forced_y1}) by the fourth-order compact finite-difference scheme leads to an inhomogeneous system of linear algebraic equations, which is solved for given roughness wavenumbers $\beta_w$ and $\alpha_w$ to obtain the shape functions of the roughness mode.

\subsection{Generalized resonance mechanisms}
When crossflow eigenmodes and roughness mode are present simultaneously in the boundary layer, interactions between them take place. Of particular interest are interactions of resonant nature, occurring at the quadratic order. These include triadic resonance and Bragg scattering generalized to unstable eigenmodes. Consider two crossflow eigenmodes with carrier waves $E_1=e^{\mathrm{i}(\alpha_1 x+\beta_1 z-\omega_1 t)}$ and $E_2=e^{\mathrm{i}(\alpha_2 x+\beta_2 z-\omega_2 t)}$, and one roughness mode $E_w=e^{\mathrm{i}(\alpha_wx+\beta_wz)}$. One of the eigenmodes interacts with the roughness mode to generate the other and vice versa, and this may take place in the two forms:
\begin{equation}\label{eq:interaction_1}
E_1=E_2*E_w
\end{equation}
and
\begin{equation}\label{eq:interaction_2}
E_1=E_2^**E_w,
\end{equation}
where the superscript $*$ denotes the complex conjugate. For (\ref{eq:interaction_1}) and (\ref{eq:interaction_2}) to hold, the corresponding wavenumbers and frequencies must satisfy the resonance conditions
\begin{equation}\label{eq:triad_resonance_difference}
\left\{\begin{array}{ll}
\Im{\alpha_1}=\Im{\alpha_2},\quad \omega_1=\omega_2, \\
\Re{\alpha_1}-\Re{\alpha_2}=\alpha_w,\quad \beta_1-\beta_2=\beta_w,
\end{array} \right.
\end{equation}
and
\begin{equation}\label{eq:triad_resonance_sum}
\left\{\begin{array}{ll}
\Im{\alpha_1}=\Im{\alpha_2},\quad \omega_1=-\omega_2, \\
\Re{\alpha_1}+\Re{\alpha_2}=\alpha_w,\quad \beta_1+\beta_2=\beta_w,
\end{array} \right.
\end{equation}
respectively.
These two forms of resonance will be referred to as triad difference and triad sum respectively. Each set of the resonance conditions has two parts: the first specifies the requirements that must be satisfied by the eigenmodes involved in the resonance, and the second determines the wavenumbers of the participating roughness mode. For the triad difference, the growth rates and frequencies of the two eigenmodes must be equal, and the roughness wavenumbers are the differences of those of the two eigenmodes; for the triad sum, the growth rates must also be equal but the frequencies must be of opposite sign, and the roughness wavenumbers are the sums of those of the two eigenmodes. 

There is a special case for the triad sum where the two eigenmodes are identical. The corresponding resonance conditions become
\begin{equation}\label{eq:bragg_scattering}
\left\{\begin{array}{ll}
\omega=0,  \\
2*\Re{\alpha}=\alpha_w,\quad 2*\beta=\beta_w.
\end{array} \right.
\end{equation}
This kind of resonance will be referred to as generalized Bragg scattering, which can take place between any stationary eigenmode and the roughness element with wavenumbers twice those of the former. 

The triad difference, triad sum and Bragg scattering are referred to as the generalized resonance mechanisms in this paper. A roughness component satisfying one of the resonant conditions (\ref{eq:triad_resonance_difference})-(\ref{eq:bragg_scattering}) is of special significance as it produces an \textit{O}($h$) correction to the growth rates of crossflow vortices. In contrast, components which do not satisfy any of (\ref{eq:triad_resonance_difference})-(\ref{eq:bragg_scattering}) interact with each eigenmode at the cubic order, namely, the roughness mode interacts with itself and the eigenmode to generate respectively an \textit{O}($h^2$) mean-flow distortion and \textit{O}($\epsilon h$) Fourier components ($\alpha_1\pm\alpha_w, \beta_1\pm\beta_w$), which interact in turn at the cubic order with the eigenmode and the roughness mode respectively, both reproducing the eigenmode. The resulting effect is a much smaller \textit{O}($h^2$) correction to the growth rate.

\begin{figure}
% \psfrag{A}{$(a)$}
% \psfrag{B}{$(b)$}
% \psfrag{x}[][]{$\beta$}
% \psfrag{y}{$\alpha_\hr$}
% \psfrag{z}{$-\alpha_\hi$}
\centerline{\includegraphics{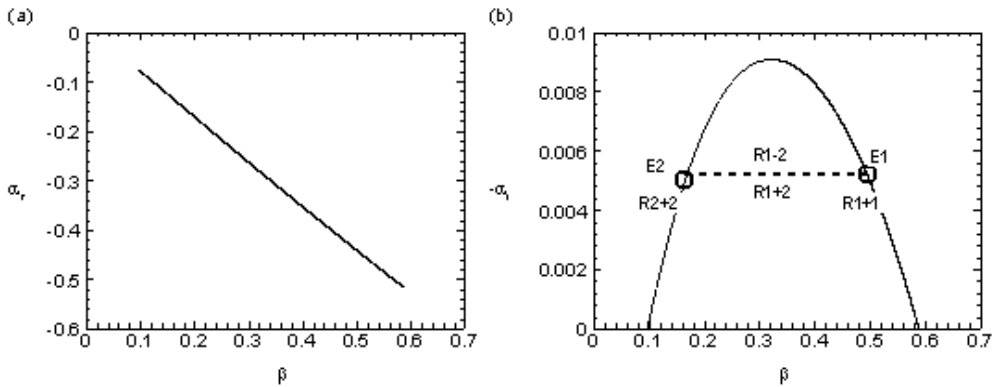}}
\caption{Instability characteristics of stationary vortices for the parameters: $\phi_0=45^\circ$, $m=0.34207$ and $Re=338$. $(a)$ The variation of $\alpha_\hr$ with $\beta$. $(b)$ The variation of $-\alpha_\hi$ with $\beta$ and the schematic of the resonances, where the letters `E' and `R' stand for eigenmode and roughness mode respectively.}
\label{fig:sv_resonance_mechanism}
\end{figure}

The crossflow eigenmodes and single-wavenumber roughness satisfying the resonance conditions of the generalized resonance mechanisms are searched for stationary and non-stationary cases by applying the first part of each set of the resonance conditions. This requires computation of unstable crossflow eigenmodes by linear stability analysis. For illustration, the results for the base-flow parameters of $\phi_0=45^\circ$, $m=0.34207$, $Re=338$ and $Re=2000$ are presented here; the value of $m$ is taken to be the same as in \citet{hogberg1998secondary} so that our linear stability results can be validated by comparison with theirs. Figure \ref{fig:sv_resonance_mechanism} displays the results for the stationary case.  As figure \ref{fig:sv_resonance_mechanism}$(a)$ shows, the variation of $\alpha_\hr$ against $\beta$ is almost linear, which is a remarkable property and will have an important implication for resonance. Figure \ref{fig:sv_resonance_mechanism}$(b)$ shows that instability exists ($-\alpha_\hi>0$) in an interval of $\beta$, the ends of which correspond to two neutral eigenmodes, both of which turn out to be relevant for understanding the response of the boundary layer to roughness. Unstable modes to the left/right of the most unstable one will be referred to as left/right branches respectively. The schematic of resonance for stationary vortices is shown in figure \ref{fig:sv_resonance_mechanism}$(b)$. Two stationary vortices can have equal growth rate ($-\alpha_\hi$), and they are denoted as E1 and E2 in the figure. They can interact with each other in the presence of suitable roughness through the triad difference and triad sum, referred to as 1-2 and 1+2, respectively. Furthermore, each mode may interact with a roughness mode through Bragg scattering, denoted by 1+1 and 2+2, respectively. Note that the crossflow eigenmodes are denoted by numbers, and the single-wavenumber roughnesses are designated by numbers and symbols `+' and `-', indicating that the resonance mechanism involved is the triad sum and difference respectively. Stationary eigenmodes 1 and 2 can also interact with a multiple-wavenumber roughness comprised of the four components, in which case multiple generalized resonance mechanisms take effect at the same time.

\begin{figure}
% \psfrag{A}{$(a)$}
% \psfrag{B}{$(b)$}
% \psfrag{x}[][]{$\beta$}
% \psfrag{y}{$\omega$}
\centerline{\includegraphics{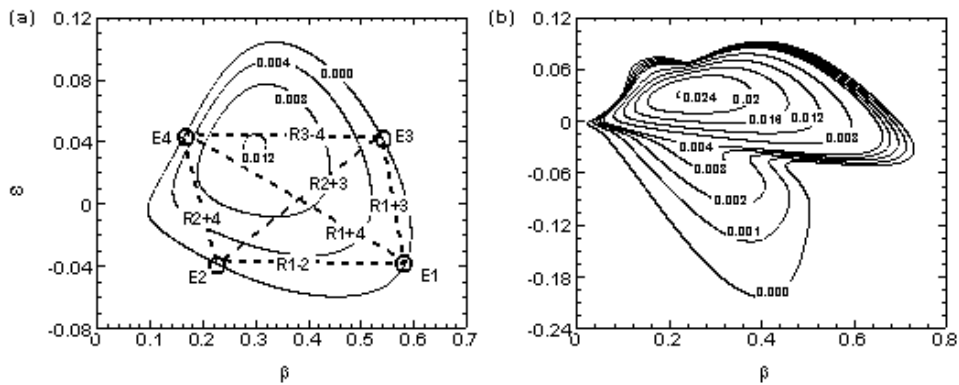}}
\caption{Contours of growth rate and schematic of the resonances for travelling vortices. The parameters are $\phi_0=45^\circ$, $m=0.34207$, $(a)$ $Re=338$ and $(b)$ $Re=2000$. The letters `E' and `R' stand for eigenmode and roughness mode respectively.}
\label{fig:tw_resonance_mechanism}
\end{figure}

Now turn to resonances involving travelling vortices. Figure \ref{fig:tw_resonance_mechanism} shows the contours of the growth rate at a low and a high Reynolds numbers. For the former, every contour line is convex so that we can locate up to four travelling eigenmodes with equal $|\omega|$ and $-\alpha_\hi$, which are labeled as E1, E2, E3 and E4 in figure \ref{fig:tw_resonance_mechanism}$(a)$. The frequency of travelling vortices 1 and 2 is negative, and that of 3 and 4 is positive. These four travelling vortices can interact with six single-wavenumber components, which are 1+3, 1+4, 2+3 and 2+4 through the triad sum, and 1-2 and 3-4 through the triad difference. Figure \ref{fig:tw_resonance_mechanism}$(b)$ indicates that at a higher $Re$ contour lines of small growth rates can be concave for certain range of $\beta$ and $\omega$, where up to six travelling eigenmodes with equal $|\omega|$ and $-\alpha_\hi$ may be located. For simplicity, in this paper we concentrate only on the case of four travelling eigenmodes, but the methodology can be easily extended to the more complex case.

There exist connections between stationary and non-stationary cases. When the frequency of the four travelling eigenmodes is reduced to zero, eigenmodes 3 and 4 coalesce with 1 and 2 respectively. It follows that roughness 3-4 becomes the same as 1-2, while roughnesses 1+4 and 2+3 collapse, both becoming equivalent to roughness 1+2 for the stationary case; see figure \ref{fig:sv_resonance_mechanism}$(b)$. Roughnesses 1+3 and 2+4 become 1+1 and 2+2, respectively, i.e. these triad-sum interactions degenerate into Bragg scattering.

\section{Mathematical description of resonant interactions between crossflow eigenmodes and roughness modes}\label{sec:derivation}
In the presence of distributed surface roughness (waviness), the evolution and even the composition of the crossflow eigenmodes will be different from the flat-surface case, due to the resonant interactions with the roughness modes at the quadratic order. In order to quantify the effects of distributed roughness, the interactions must be analysed. A large-Reynolds-number asymptotic analysis is performed by \citet{butler2018vortex} for stationary vortices. In the present study, we will take a finite-Reynolds-number approach, and consider both travelling and stationary vortices.
Using a multi-scale method, we will derive the amplitude equations for the eigenmodes, first for their interactions with a single-wavenumber roughness. The analysis will then be extended to the case of multiple-wavenumber roughness.
The quadratic interactions cause an \textit{O}$(h)$ correction to the growth rate. In order to account for this effect, we introduce a slow variable $\bar{x}=hx$, and let the amplitudes of the eigenmodes be functions of this slow variable.

\subsection{Interaction of eigenmodes with single-wavenumber roughness: triad difference}
The interaction through the triad difference involves two eigenmodes $j$ and $k$, and one single-wavenumber roughness.  The disturbance $\phi$ (standing for $u$, $v$, $w$ and $p$), comprising of the roughness mode and the eigenmodes, can be represented by
\begin{equation}\label{eq:disturbance_triad_difference}
\phi=h[\hat{\phi}_wE_w+c.c.]+\epsilon[A_j(\bar{x})\hat{\phi}_{j0}(y)E_j+h\hat{\phi}_{j1}E_j+c.c.]+\epsilon[A_k(\bar{x})\hat{\phi}_{k0}(y)E_k+h\hat{\phi}_{k1}E_k+c.c.],
\end{equation}
where $E_w=e^{\mathrm{i}(\alpha_wx+\beta_wz)}$, $E_j=e^{\mathrm{i}(\alpha_jx+\beta_jz-\omega_jt)}$ and $E_k=e^{\mathrm{i}(\alpha_kx+\beta_kz-\omega_kt)}$. $A_j$ and $A_k$ denote the amplitude functions of the respective eigenmodes. Substituting (\ref{eq:disturbance_triad_difference}) into (\ref{eq:nonlinear_disturbance_equations}), at \textit{O}($\epsilon$) and \textit{O}($h$) we recover respectively the eigenvalue problem and the inhomogeneous problem determining the roughness mode. At \textit{O}($\epsilon h$), we obtain the equations for $\hat{\phi}_{j1}$ and $\hat{\phi}_{k1}$
\begin{subeqnarray}\label{eq:second_order_equations_triad_difference}
\frac{\mathrm{d} \hat{\boldsymbol{\varphi}}_{j1}}{\mathrm{d}y}=\mathsfbi{L}_{j}\hat{\boldsymbol{\varphi}}_{j1}+\frac{\mathrm{d}A_j}{\mathrm{d}\bar{x}}\boldsymbol{f}_{j}+A_k\boldsymbol{f}_{k-w},\\  
\frac{\mathrm{d} \hat{\boldsymbol{\varphi}}_{k1}}{\mathrm{d}y}=\mathsfbi{L}_{k}\hat{\boldsymbol{\varphi}}_{k1}+\frac{\mathrm{d}A_k}{\mathrm{d}\bar{x}}\boldsymbol{f}_{k}+A_j\boldsymbol{f}_{j-w^*},
\end{subeqnarray}
where $\hat{\boldsymbol{\varphi}}_{j,k1}$ and $\mathsfbi{L}_{j,k}$ have the same expressions as $\hat{\boldsymbol{\varphi}}$ and $\mathsfbi{L}$ in (\ref{eq:one_order_OS})-(\ref{eq:matrix}) provided that $\alpha$ and $\beta$ are replaced by $\alpha_{j,k}$ and $\beta_{j,k}$ respectively. The inhomogeneous terms are
\begin{eqnarray}
\boldsymbol{f}_{j}&=&[0,-f_{j}^c,0,Ref_{j}^x,-Df_{j}^c/Re-f_{j}^y,Ref_{j}^z]^T,\nonumber \\ 
% \boldsymbol{f}_{k}&=&[0,f_{k}^c,0,-Re*f_{k}^x,Df_{k}^c/Re+f_{k}^y,-Re*f_{k}^z]^T,\nonumber \\
\boldsymbol{f}_{k-w}&=&[0,0,0,Ref_{k-w}^x,-f_{k-w}^y,Ref_{k-w}^z]^T,\nonumber 
% \boldsymbol{f}_{j-w^*}&=&[0,0,0,-Re*f_{j-w^*}^x,f_{j-w^*}^y,-Re*f_{j-w^*}^z]^T,\nonumber
\end{eqnarray}
where
\begin{eqnarray}
f_j^x=(U-\frac{2\mathrm{i}\alpha_j}{Re})\hat{u}_{j0}+\hat{p}_{j0},f_j^y=(U-\frac{2\mathrm{i}\alpha_j}{Re})\hat{v}_{j0},f_j^z=(U-\frac{2\mathrm{i}\alpha_j}{Re})\hat{w}_{j0},f_j^c=\hat{u}_{j0},\nonumber
% f_k^x=-[(U-\frac{2\mathrm{i}\alpha_k}{Re})\hat{u}_{k0}+\hat{p}_{k0}],f_k^y=-[(U-\frac{2\mathrm{i}\alpha_k}{Re})\hat{v}_{k0}],f_k^z=-[(U-\frac{2\mathrm{i}\alpha_k}{Re})\hat{w}_{k0}],f_k^c=-\hat{u}_{k0},\nonumber
\end{eqnarray}
and the Reynolds stresses
\begin{eqnarray}
f_{k-w}^x&=&\mathrm{i}\alpha_w\hat{u}_{k0}\hat{u}_w+\mathrm{i}\alpha_k\hat{u}_{k0}\hat{u}_w+\hat{v}_{k0}D\hat{u}_w+D\hat{u}_{k0}\hat{v}_w+\mathrm{i}\beta_w\hat{w}_{k0}\hat{u}_w+\mathrm{i}\beta_k\hat{u}_{k0}\hat{w}_w,\nonumber\\[3pt]
f_{k-w}^y&=&\mathrm{i}\alpha_w\hat{u}_{k0}\hat{v}_w+\mathrm{i}\alpha_k\hat{v}_{k0}\hat{u}_w+\hat{v}_{k0}D\hat{v}_w+D\hat{v}_{k0}\hat{v}_w+\mathrm{i}\beta_w\hat{w}_{k0}\hat{v}_w+\mathrm{i}\beta_k\hat{v}_{k0}\hat{w}_w,\nonumber\\[3pt]
f_{k-w}^z&=&\mathrm{i}\alpha_w\hat{u}_{k0}\hat{w}_w+\mathrm{i}\alpha_k\hat{w}_{k0}\hat{u}_w+\hat{v}_{k0}D\hat{w}_w+D\hat{w}_{k0}\hat{v}_w+\mathrm{i}\beta_w\hat{w}_{k0}\hat{w}_w+\mathrm{i}\beta_k\hat{w}_{k0}\hat{w}_w.\nonumber
% f_{j-w^*}^x&=&-(-\mathrm{i}\alpha_w\hat{u}_{j0}\hat{u}_w^*+\mathrm{i}\alpha_j\hat{u}_{j0}\hat{u}_w^*+\hat{v}_{j0}D\hat{u}_w^*+D\hat{u}_{j0}\hat{v}_w^*-\mathrm{i}\beta_w\hat{w}_{j0}\hat{u}_w^*+\mathrm{i}\beta_j\hat{u}_{j0}\hat{w}_w^*),\nonumber\\[3pt]
% f_{j-w^*}^y&=&-(-\mathrm{i}\alpha_w\hat{u}_{j0}\hat{v}_w^*+\mathrm{i}\alpha_j\hat{v}_{j0}\hat{u}_w^*+\hat{v}_{j0}D\hat{v}_w^*+D\hat{v}_{j0}\hat{v}_w^*-\mathrm{i}\beta_w\hat{w}_{j0}\hat{v}_w^*+\mathrm{i}\beta_j\hat{v}_{j0}\hat{w}_w^*),\nonumber\\[3pt]
% f_{j-w^*}^z&=&-(-\mathrm{i}\alpha_w\hat{u}_{j0}\hat{w}_w^*+\mathrm{i}\alpha_j\hat{w}_{j0}\hat{u}_w^*+\hat{v}_{j0}D\hat{w}_w^*+D\hat{w}_{j0}\hat{v}_w^*-\mathrm{i}\beta_w\hat{w}_{j0}\hat{w}_w^*+\mathrm{i}\beta_j\hat{w}_{j0}\hat{w}_w^*).\nonumber
\end{eqnarray}
The expression for $\boldsymbol{f}_{k}$ follows from $\boldsymbol{f}_{j}$ by changing $j$ to $k$, and the expression for $\boldsymbol{f}_{j-w^*}$ from $\boldsymbol{f}_{k-w}$ by the substitutions:
\begin{eqnarray}
k \to j, \quad (\alpha_w, \beta_w)\to(-\alpha_w, -\beta_w) \quad \mbox{and\ } \quad \phi_w\to\phi_w^*.\nonumber
\end{eqnarray}
The corresponding boundary conditions at $y=0$ for (\ref{eq:second_order_equations_triad_difference}) are given by
\begin{subeqnarray}\label{eq:boundary_conditions_triad_difference}
\hat{u}_{j1}(0)=-A_k\hat{u}_{k0}'(0), \quad \hat{v}_{j1}(0)=0, \quad \hat{w}_{j1}(0)=-A_k\hat{w}_{k0}'(0); \\
\hat{u}_{k1}(0)=-A_j\hat{u}_{j0}'(0), \quad \hat{v}_{k1}(0)=0, \quad \hat{w}_{k1}(0)=-A_j\hat{w}_{j0}'(0).
\end{subeqnarray}
The far-field boundary conditions are homogeneous as usual.

In order for the inhomogeneous boundary-value problem (\ref{eq:second_order_equations_triad_difference}$a$) with (\ref{eq:boundary_conditions_triad_difference}$a$), or (\ref{eq:second_order_equations_triad_difference}$b$) with (\ref{eq:boundary_conditions_triad_difference}$b$), to have a solution, a solvability condition must be satisfied. This can be derived by reducing (\ref{eq:second_order_equations_triad_difference}$a$) or (\ref{eq:second_order_equations_triad_difference}$b$) to an inhomogeneous O-S equation, and then using the adjoint eigenfunction. Alternatively, we may discretize (\ref{eq:second_order_equations_triad_difference}$a$) and (\ref{eq:second_order_equations_triad_difference}$b$), and apply Gaussian elimination to the resulting algebraic systems. Either way, we obtain two coupled amplitude equations,
\begin{subeqnarray}\label{eq:triad_difference_amplitude_equations}
\frac{\mathrm{d}A_j}{\mathrm{d}\bar{x}}+F_{k-w}A_k(\bar{x})=0,\\
\frac{\mathrm{d}A_k}{\mathrm{d}\bar{x}}+F_{j-w^*}A_j(\bar{x})=0,
\end{subeqnarray}
where $F_{k-w}$ and $F_{j-w^*}$ are constants calculated by applying the solvability condition; the two procedures mentioned above give the same value for each of these constants. The above amplitude equations can be rearranged into the matrix form
\begin{equation}
\setlength{\arraycolsep}{0pt}
\renewcommand{\arraystretch}{1.3}
\frac{\mathrm{d}}{\mathrm{d}\bar{x}}
[A_j, A_k]^T
=
\mathsfbi{F}
[A_j, A_k]^T
\end{equation}
with $\mathsfbi{F}$ being the $2\times2$ matrix,
\begin{equation}
\setlength{\arraycolsep}{0pt}
\renewcommand{\arraystretch}{1.3}
 \mathsfbi{F}=-
 \left[
\begin{array}{cc}
 0&F_{k-w}\\
 F_{j-w^*}&0
\end{array}  \right]. \nonumber
\end{equation}
Seeking solutions of the exponential form,
\begin{equation}
[A_j,A_k]^T=\boldsymbol{a}e^{\lambda\bar{x}}=[a_j,a_k]^Te^{\lambda\bar{x}},
\end{equation}
leads to
an eigenvalue problem
\begin{equation}
\lambda\boldsymbol{a}=\mathsfbi{F}\boldsymbol{a}.
\end{equation}

The physical meanings of the eigenvalue $\lambda=\lambda_\hr+\hi\lambda_\hi$ and eigenvector $\boldsymbol{a}$ are now analyzed. In terms of $\boldsymbol{a}$ and $\lambda$, the chordwise disturbance velocity $u_j$ and $u_k$ at the leading order can be written as
\begin{subeqnarray}
A_j\hat{u}_{j0}E_j=a_j\hat{u}_{j0}e^{(\lambda_\hr h-\Im\alpha_j)x}e^{\mathrm{i}(\Re\alpha_jx+\lambda_\hi hx+\beta_jz-\omega_jt)},\\
A_k\hat{u}_{k0}E_k=a_k\hat{u}_{k0}e^{(\lambda_\hr h-\Im\alpha_k)x}e^{\mathrm{i}(\Re\alpha_kx+\lambda_\hi hx+\beta_kz-\omega_kt)},
\end{subeqnarray}
indicating that the new growth rates are $(\lambda_\hr h-\Im\alpha_j)$ for eigenmode $j$ and $(\lambda_\hr h-\Im\alpha_k)$ for eigenmode $k$ after the resonant interaction. The growth rates of both eigenmodes $j$ and $k$ are modified by $\lambda_\hr h$, which is proportional to the roughness height $h$. We define the scaling factor as the growth-rate correction coefficient, which is dependent only on the resonant triad. The latter is uniquely determined by a pair of $(-\alpha_\hi, \omega)$, and the growth-rate correction coefficient $\lambda_\hr$  is therefore  a function of $-\alpha_\hi$ and $\omega$,
\begin{equation}
\lambda_\hr=\lambda_\hr(-\alpha_\hi, \omega).
\end{equation}

The eigenvector $\boldsymbol{a}$ represents the ratio of the amplitudes of the two interacting eigenmodes.
The amplitude ratio, defined as
\begin{equation}
\rho(-\alpha_\hi, \omega)=|{a_j}/{a_k}|,
\end{equation}
is also a function of $-\alpha_\hi$ and $\omega$. It transpires that the two eigenmodes are coupled  by small-amplitude roughness, whereas they are independent of each other for the smooth-wall case in the linear regime.

\subsection{Interaction of eigenmodes with single-wavenumber roughness: triad sum}
The disturbance $\phi$ remains of the form (\ref{eq:disturbance_triad_difference}).
% \begin{equation}\label{eq:disturbance_triad_sum}
% \phi=h[\hat{\phi}_wE_w+c.c.]+\epsilon[A_j(\bar{x})\hat{\phi}_{j0}(y)E_j+h*\hat{\phi}_{j1}E_j+c.c.]+\epsilon[A_k(\bar{x})\hat{\phi}_{k0}(y)E_k+h*\hat{\phi}_{k1}E_k+c.c.]. \nonumber
% \end{equation}
By a similar procedure as that for the triad difference, the equations at \textit{O}($\epsilon h$) are given as
\begin{subeqnarray}\label{eq:second_order_equations_triad_sum}
\frac{\mathrm{d} \hat{\boldsymbol{\varphi}}_{j1}}{\mathrm{d}y}=\mathsfbi{L}_{j0}\hat{\boldsymbol{\varphi}}_{j1}+\frac{\mathrm{d}A_j}{\mathrm{d}\bar{x}}\boldsymbol{f}_j+A_k^*\boldsymbol{f}_{k^*-w},\\
\frac{\mathrm{d} \hat{\boldsymbol{\varphi}}_{k1}}{\mathrm{d}y}=\mathsfbi{L}_{k0}\hat{\boldsymbol{\varphi}}_{k1}+\frac{\mathrm{d}A_k}{\mathrm{d}\bar{x}}\boldsymbol{f}_k+A_j^*\boldsymbol{f}_{j^*-w},
\end{subeqnarray}
where $\boldsymbol{f}_j$ and $\boldsymbol{f}_k$ have the same expressions as those for the triad difference, and $\boldsymbol{f}_{k^*-w}$ is given by
\begin{equation}
\boldsymbol{f}_{k^*-w}=[0,0,0,Ref_{k^*-w}^x,-f_{k^*-w}^y,Ref_{k^*-w}^z]^T, \nonumber 
% \boldsymbol{f}_{j^*-w}&=&[0,0,0,-Re*f_{j^*-w}^x,f_{j^*-w}^y,-Re*f_{j^*-w}^z]^T   \nonumber
\end{equation}
with the Reynolds stresses
\begin{eqnarray}
f_{k^*-w}^x&=&\mathrm{i}\alpha_{w}\hat{u}_{k0}^*\hat{u}_{w}-\mathrm{i}\alpha_k^*\hat{u}_{k0}^*\hat{u}_{w}+\hat{v}_{k0}^*D\hat{u}_{w}+D\hat{u}_{k0}^*\hat{v}_{w}+\mathrm{i}\beta_{w}\hat{w}_{k0}^*\hat{u}_{w}-\mathrm{i}\beta_k\hat{u}_{k0}^*\hat{w}_{w},\nonumber\\[3pt]
f_{k^*-w}^y&=&\mathrm{i}\alpha_{w}\hat{u}_{k0}^*\hat{v}_{w}-\mathrm{i}\alpha_k^*\hat{v}_{k0}^*\hat{u}_{w}+\hat{v}_{k0}^*D\hat{v}_{w}+D\hat{v}_{k0}^*\hat{v}_{w}+\mathrm{i}\beta_{w}\hat{w}_{k0}^*\hat{v}_{w}-\mathrm{i}\beta_k\hat{v}_{k0}^*\hat{w}_{w},\nonumber\\[3pt]
f_{k^*-w}^z&=&\mathrm{i}\alpha_{w}\hat{u}_{k0}^*\hat{w}_{w}-\mathrm{i}\alpha_k^*\hat{w}_{k0}^*\hat{u}_{w}+\hat{v}_{k0}^*D\hat{w}_{w}+D\hat{w}_{k0}^*\hat{v}_{w}+\mathrm{i}\beta_{w}\hat{w}_{k0}^*\hat{w}_{w}-\mathrm{i}\beta_k\hat{w}_{k0}^*\hat{w}_{w}.\nonumber
% f_{j^*-w}^x&=&-(\mathrm{i}\alpha_{w}\hat{u}_{j0}^*\hat{u}_{w}-\mathrm{i}\alpha_j^*\hat{u}_{j0}^*\hat{u}_{w}+\hat{v}_{j0}^*D\hat{u}_{w}+D\hat{u}_{j0}^*\hat{v}_{w}+\mathrm{i}\beta_{w}\hat{w}_{j0}^*\hat{u}_{w}-\mathrm{i}\beta_j\hat{u}_{j0}^*\hat{w}_{w}),\nonumber \\[3pt]
% f_{j^*-w}^y&=&-(\mathrm{i}\alpha_{w}\hat{u}_{j0}^*\hat{v}_{w}-\mathrm{i}\alpha_j^*\hat{v}_{j0}^*\hat{u}_{w}+\hat{v}_{j0}^*D\hat{v}_{w}+D\hat{v}_{j0}^*\hat{v}_{w}+\mathrm{i}\beta_{w}\hat{w}_{j0}^*\hat{v}_{w}-\mathrm{i}\beta_j\hat{v}_{j0}^*\hat{w}_{w}),\nonumber \\[3pt]
% f_{j^*-w}^z&=&-(\mathrm{i}\alpha_{w}\hat{u}_{j0}^*\hat{w}_{w}-\mathrm{i}\alpha_j^*\hat{w}_{j0}^*\hat{u}_{w}+\hat{v}_{j0}^*D\hat{w}_{w}+D\hat{w}_{j0}^*\hat{v}_{w}+\mathrm{i}\beta_{w}\hat{w}_{j0}^*\hat{w}_{w}-\mathrm{i}\beta_j\hat{w}_{j0}^*\hat{w}_{w}).\nonumber
\end{eqnarray}
The expression for $\boldsymbol{f}_{j^*-w}$ follows from $\boldsymbol{f}_{k^*-w}$ by replacing $k$ by $j$.
The corresponding boundary conditions at $y=0$ are
\begin{subeqnarray}\label{eq:boundary_conditions_triad_sum}
\hat{u}_{j1}=-A_k^*\hat{u}_{k0}^{*'}(0), \quad \hat{v}_{j1}=0, \quad \hat{w}_{j1}=-A_k^*\hat{w}_{k0}^{*'}(0); \\
\hat{u}_{k1}=-A_j^*\hat{u}_{j0}^{*'}(0), \quad \hat{v}_{k1}=0, \quad \hat{w}_{k1}=-A_j^*\hat{w}_{j0}^{*'}(0),
\end{subeqnarray}
and the boundary conditions far away from the wall are homogeneous as usual.

Applying solvability conditions to (\ref{eq:second_order_equations_triad_sum}$a$) with (\ref{eq:boundary_conditions_triad_sum}$a$) and (\ref{eq:second_order_equations_triad_sum}$b$) with (\ref{eq:boundary_conditions_triad_sum}$b$) leads to two coupled
 amplitude equations
\begin{subeqnarray}\label{eq:amplitude_equations_triad_sum}
\frac{\mathrm{d}A_j}{\mathrm{d}\bar{x}}+F_{k^*-w}A_k^*=0, \\
\frac{\mathrm{d}A_k}{\mathrm{d}\bar{x}}+F_{j^*-w}A_j^*=0.
\end{subeqnarray}
Note that the above two complex equations are somewhat different from (\ref{eq:triad_difference_amplitude_equations}$a$)-(\ref{eq:triad_difference_amplitude_equations}$b$) in that they involve complex conjugates of the amplitude functions. Decoupling them into four real equations, we can recast the system into the matrix form,
\begin{equation}\label{eq:triad_sum_matrix_equation}
\setlength{\arraycolsep}{0pt}
\renewcommand{\arraystretch}{1.3}
\frac{\mathrm{d}}{\mathrm{d}\bar{x}}
[\Re{A}_j, \Im{A}_j, \Re{A}_k, \Im{A}_k]^T
=
\mathsfbi{F}
[\Re{A}_j, \Im{A}_j, \Re{A}_k, \Im{A}_k]^T,
\end{equation}
where the $4\times4$ coefficient matrix 
\begin{equation}\label{eq:triad_sum_coe_matrix}
\setlength{\arraycolsep}{0pt}
\renewcommand{\arraystretch}{1.3}
 \mathsfbi{F}=-
 \left[
\begin{array}{cccc}
 0\quad&0\quad&\Re F_{k^*-w}\quad& \Im F_{k^*-w}\\
 0\quad&0\quad&\Im F_{k^*-w}\quad&-\Re F_{k^*-w}\\
    \Re F_{j^*-w}\quad&\Im F_{j^*-w}\quad&0\quad&0\\
 \Im F_{j^*-w}\quad&-\Re F_{j^*-w}\quad&0\quad&0
\end{array}  \right]
\end{equation}
is real-valued. Equation (\ref{eq:triad_sum_matrix_equation}) admits solutions of the exponential form,
\begin{equation}\label{eq:triad_sum_exponential_solution}
[\Re{A}_j, \Im{A}_j, \Re{A}_k, \Im{A}_k]^T=\frac{1}{2}(\boldsymbol{a}e^{\lambda\bar{x}}+c.c.),
\end{equation}
substituting of which into (\ref{eq:triad_sum_matrix_equation}) yields the eigenvalue problem, $\lambda\boldsymbol{a}=\mathsfbi{F}\boldsymbol{a}$. The real part of $\lambda$, $\lambda_\hr$, represents the growth-rate correction coefficient, the same as that for the triad difference. However, the definition of the amplitude ratio $\rho$ for the triad sum is different. Note that the eigenvector $\boldsymbol{a}$ can be expressed in the polar form,
\begin{equation}\label{eq:polar_form}
\boldsymbol{a}=\left[a_{j1}, a_{j2}, a_{k1}, a_{k2}\right]^T=\left[|a_{j1}|e^{\hi\theta_{j1}}, |a_{j2}|e^{\hi\theta_{j2}}, |a_{k1}|e^{\hi\theta_{k1}}, |a_{k2}|e^{\hi\theta_{k2}}\right]^T.
\end{equation}
 According to (\ref{eq:triad_sum_matrix_equation}), (\ref{eq:triad_sum_exponential_solution}) and (\ref{eq:polar_form}), the modulus of $A_{j,k}$ can be expressed as 
\begin{equation}\label{eq:modulus_A}
|A_{j,k}|^2=e^{2\lambda_\hr\bar{x}}|a_{j,k}|^2=e^{2\lambda_\hr\bar{x}}\left[|a_{{j,k}1}|^2\cos^2(\lambda_\hi\bar{x}+\theta_{{j,k}1})+|a_{{j,k}2}|^2\cos^2(\lambda_\hi\bar{x}+\theta_{{j,k}2})\right]. 
\end{equation}
% where $\lambda_\hi$ is the imaginary part of the eigenvalue $\lambda$. 
For the triad sum, with the coefficient matrix $\mathsfbi{F}$ given by (\ref{eq:triad_sum_coe_matrix}), the elements of the eigenvector $\boldsymbol{a}$ satisfy
\begin{equation}\label{eq:elements_of_eigenvector}
a_{{j}1}=\hi a_{{j}2}, \quad a_{{k}1}=\hi a_{{k}2}, \nonumber
\end{equation}
from which it follows $|a_{{j}1}|=|a_{{j}2}|$,  $|a_{{k}1}|=|a_{{k}2}|$, $\theta_{{j}1}=\theta_{{j}2}+\pi/2$ and $\theta_{{k}1}=\theta_{{k}2}+\pi/2$.
The amplitude ratio $\rho$ for the triad sum is thus defined as 
\begin{equation}\label{eq:triad_sum_amplitude_ratio}
\rho(-\alpha_\hi, \omega)=|a_{j1}|/|a_{k1}|.
\end{equation}
For the interaction between the eigenmodes and multiple-wavenumber roughness, which will be discussed later, the relations in (\ref{eq:elements_of_eigenvector}) may not hold.  Using trigonometric identities, $a_{j,k}$ can be expressed as
\begin{equation}\label{eq:a_jk_trigonometric_form}
|a_{j,k}|^2=G_{j,k}\cos(2\lambda_\hi\bar{x})-H_{j,k}\sin(2\lambda_\hi\bar{x})+(|a_{{j,k}1}|^2+|a_{{j,k}2}|^2)/2,
\end{equation}
where $G_{j,k}=|a_{{j,k}1}|^2\cos(2\theta_{{j,k}1})/2+|a_{{j,k}2}|^2\cos(2\theta_{{j,k}2})/2$ and $H_{j,k}=|a_{{j,k}1}|^2\sin(2\theta_{{j,k}1})/2+|a_{{j,k}2}|^2\sin(2\theta_{{j,k}2})/2$. The maximum value of $|a_{j,k}|$ is given by
\begin{equation}
|a_{j,k}|^2_{\mathrm{max}}=\sqrt{G_{j,k}^2+H_{j,k}^2}+(|a_{{j,k}1}|^2+|a_{{j,k}2}|^2)/2.
\end{equation}
% where we can recover (\ref{eq:a_jk}) with $G_{j,k}=H_{j,k}=0$ for the triad-sum interaction. 
Now the amplitude ratio is defined as
\begin{equation}\label{eq:amplitude_ratio_multiple}
\rho=|a_j|_{\mathrm{max}}/|a_k|_{\mathrm{max}},
\end{equation}
from which we can recover (\ref{eq:triad_sum_amplitude_ratio}) with $|a_{j,k}|_{\mathrm{max}}=|a_{{j,k}1}|$ following from the relations in (\ref{eq:elements_of_eigenvector}).
Note that for the triad sum the frequencies of the two eigenmodes are of opposite sign. We use the positive one as the control parameter for $\lambda_\hr$ and $\rho$.

For stationary vortices, the triad difference and sum both operate, but an additional interaction, Bragg scattering, exists. Following a similar procedure as for the triad sum, the amplitude equation can be derived as  
\begin{equation}
\frac{\mathrm{d}A}{\mathrm{d}\bar{x}}+F_{w}A^*=0,
\end{equation}
which can be rewritten into a $2\times2$ system for $[\Re A, \Im A]^T$.
The growth-rate correction coefficient for Bragg scattering can be obtained by the same method as that for the triad sum. Since only one stationary eigenmode is involved in Bragg scattering, the issue of amplitude ratio does not arise. 

\subsection{Interaction of crossflow eigenmodes with multiple-wavenumber roughness}
The analysis is now extended to the interactions between crossflow eigenmodes and multiple-wavenumber roughness. Stationary and non-stationary cases must be considered separately due to different forms of resonance.

Two stationary vortices with the same growth rate can in general interact simultaneously with four roughness modes, namely, the triad sum and difference interactions with $(\alpha_{w_{1\pm2}}, \beta_{w_{1\pm2}})$, and Bragg scattering by $(\alpha_{w_{1+1}}, \beta_{w_{1+1}})$ and $(\alpha_{w_{2+2}}, \beta_{w_{2+2}})$. The disturbance comprised of these modes takes the form,
\begin{eqnarray}
\phi=&h&(\hat{\phi}_{w_{1-2}}E_{w_{1-2}}+c.c.)+h(\hat{\phi}_{w_{1+2}}E_{w_{1+2}}+c.c.) \nonumber \\
&+&h(\hat{\phi}_{w_{1+1}}E_{w_{1+1}}+c.c.)+h(\hat{\phi}_{w_{2+2}}E_{w_{2+2}}+c.c.) \nonumber   \\
&+&\epsilon[A_1(\bar{x})\hat{\phi}_{10}(y)E_1+h\hat{\phi}_{11}E_1+c.c.]+\epsilon[A_2(\bar{x})\hat{\phi}_{20}(y)E_2+h\hat{\phi}_{21}E_2+c.c.]. \nonumber
\end{eqnarray}
Following similar steps as before, we can derive the two coupled amplitude equations,
\begin{subeqnarray}
\frac{\mathrm{d}A_1}{\mathrm{d}\bar{x}}&+\mathcal{F}_{2-w_{1-2}}A_2(\bar{x})+\mathcal{F}_{2^*-w_{1+2}}A_2^*(\bar{x})+\mathcal{F}_{1^*-w_{1+1}}A_1^*(\bar{x})&=0,\\[3pt]
\frac{\mathrm{d}A_2}{\mathrm{d}\bar{x}}&+\mathcal{F}_{1-w_{1-2}^*}A_1(\bar{x})+\mathcal{F}_{1^*-w_{1+2}}A_1^*(\bar{x})+\mathcal{F}_{2^*-w_{2+2}}A_2^*(\bar{x})&=0.
\end{subeqnarray}
The above two complex equations are again recast into a system of four real equations akin to (\ref{eq:triad_sum_matrix_equation}). Seeking exponential solutions to the system, we obtain an eigenvalue problem, $\lambda\boldsymbol{a}=\mathsfbi{F}\boldsymbol{a}$, with $\mathsfbi{F}$ being a $4\times4$ real-valued matrix.

For travelling vortices, resonances can take place among four travelling eigenmodes (with equal $\alpha_\hi$ and $|\omega|$) and six roughness modes. The disturbance consisting of these components takes the form,
\begin{eqnarray}
\phi=&h&(\hat{\phi}_{w_{1-2}}E_{w_{1-2}}+c.c.)+h(\hat{\phi}_{w_{1+3}}E_{w_{1+3}}+c.c.)+h(\hat{\phi}_{w_{1+4}}E_{w_{1+4}}+c.c.) \nonumber \\
&+&h(\hat{\phi}_{w_{2+3}}E_{w_{2+3}}+c.c.)+h(\hat{\phi}_{w_{2+4}}E_{w_{2+4}}+c.c.)+h(\hat{\phi}_{w_{3-4}}E_{w_{3-4}}+c.c.) \nonumber \\
&+&\epsilon[A_1(\bar{x})\hat{\phi}_{10}(y)E_1+h\hat{\phi}_{11}E_1+c.c.]+\epsilon[A_2(\bar{x})\hat{\phi}_{20}(y)E_2+h\hat{\phi}_{21}E_2+c.c.] \nonumber \\
&+&\epsilon[A_3(\bar{x})\hat{\phi}_{30}(y)E_3+h\hat{\phi}_{31}E_3+c.c.]+\epsilon[A_4(\bar{x})\hat{\phi}_{40}(y)E_4+h\hat{\phi}_{41}E_4+c.c.]. \nonumber
\end{eqnarray}
Following similar steps as before, we can derive the four coupled amplitude equations,
\begin{subeqnarray}
\frac{\mathrm{d}A_1}{\mathrm{d}\bar{x}}+&F_{2-w_{1-2}}A_2    +F_{3^*-w_{1+3}}A_3^*+F_{4^*-w_{1+4}}A_4^*=&0,\\[3pt]
\frac{\mathrm{d}A_2}{\mathrm{d}\bar{x}}+&F_{1-w_{1-2}^*}A_1+F_{3^*-w_{2+3}}A_3^*+F_{4^*-w_{2+4}}A_4^*=&0,\\[3pt]
\frac{\mathrm{d}A_3}{\mathrm{d}\bar{x}}+&F_{1^*-w_{1+3}}A_1^*+F_{2^*-w_{2+3}}A_2^*+F_{4-w_{3-4}}A_4    =&0,\\[3pt]
\frac{\mathrm{d}A_4}{\mathrm{d}\bar{x}}+&F_{1^*-w_{1+4}}A_1^*+F_{2^*-w_{2+4}}A_2^*+F_{3-w_{3-4}^*}A_3=&0.
\end{subeqnarray}
The above four complex equations need to be decomposed into a system of eight real equations. Seeking exponential solutions leads to an eigenvalue problem, $\lambda\boldsymbol{a}=\mathsfbi{F}\boldsymbol{a}$, with $\mathsfbi{F}$ being a $8\times8$ real-valued matrix.

For multiple-wavenumber cases, the growth-rate correction coefficient $\lambda_\hr$ can be defined as the largest real part of the eigenvalues, and the amplitude ratio $\rho$ is computed by a procedure described earlier; see (\ref{eq:a_jk_trigonometric_form})-(\ref{eq:amplitude_ratio_multiple}).

\section{Numerical results}\label{sec:results}
We now compute the growth-rate correction coefficient and amplitude ratio for representative base-flow parameters. The acceleration parameter $m$ is set to be 0.34207, 0.5 and 1.0. A local sweep angle of $45^\circ$ and two different Reynolds numbers (338 and 2000) are chosen. 

\subsection{Resonant interactions of stationary vortices with roughness modes}

\begin{figure}
% \psfrag{x}[][]{$-\alpha_\hi$}
% \psfrag{y}{$\beta_w$}
% \psfrag{z}{$\alpha_w$}
% \psfrag{A}{$(a)$}
% \psfrag{B}{$(b)$}
\centerline{\includegraphics[scale=0.9]{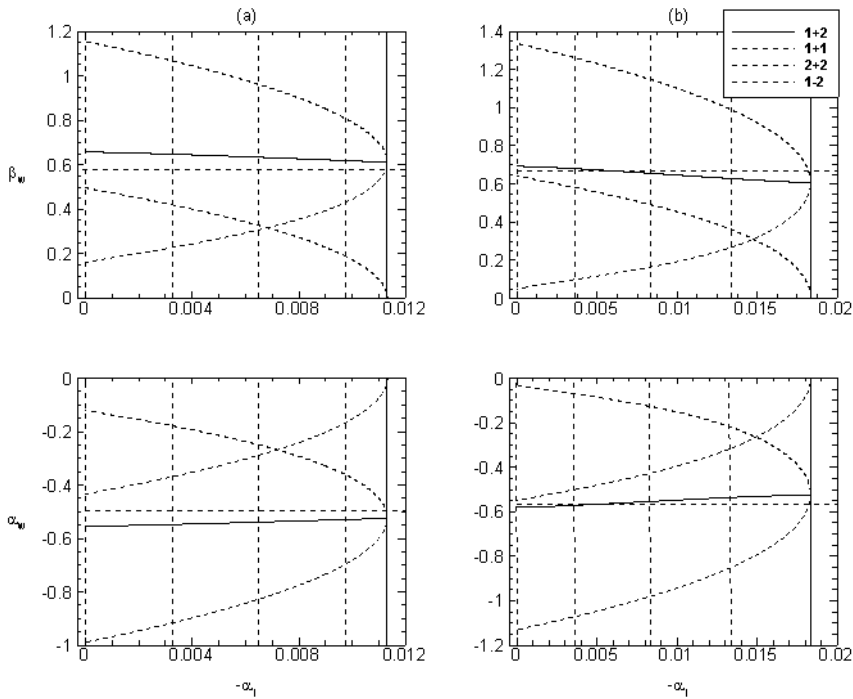}}
\caption{Variations of the roughness wavenumbers satisfying the resonance conditions of the triad sum (1+2), triad difference (1-2) and Bragg scattering (1+1 and 2+2) with the growth rate for stationary vortices at $m=0.5$. The long dashed lines stand for $\beta_s^r$ and $\alpha_s^r$, the wavenumbers of the right-branch neutral stationary eigenmodes. $(a)$ $Re=338$; $(b)$ $Re=2000$.}
\label{fig:sv_roughness_wavenumber}
\end{figure}

\begin{table}
\begin{subtable}{1\textwidth}
\centering
  \caption{$(a)\quad Re=338$}
  \begin{tabular}{lccccc}
            & $\alpha_\hi$ \quad\quad & $\beta_1$ & $\Re{\alpha_1}$  \quad\quad & $\beta_2$ & $\Re{\alpha_2}$       \\[3pt]
   case 1   & -0.0113134 \quad\quad & 0.3113300 & -0.2673790  \quad \quad& 0.2999562 & -0.2572778          \\
   case 2   & -0.0097570 \quad\quad & 0.4037300 & -0.3482950  \quad\quad & 0.2150391 & -0.1809680         \\
   case 3   & -0.0065036 \quad\quad & 0.4805300 & -0.4139413  \quad\quad & 0.1533736 & -0.1248766         \\ 
   case 4   & -0.0032529 \quad\quad & 0.5337300 & -0.4585129 \quad \quad & 0.1138362 & -0.0891467         \\ 
   case 5   &  0.0000000 \quad\quad & 0.5772319 & -0.4943765  \quad\quad & 0.0809225 & -0.0603090         
  \end{tabular}
\end{subtable}
  \\[10pt]
  \begin{subtable}{1\textwidth}
  \centering
  \caption{$(b)\quad Re=2000$}
\begin{tabular}{lccccc}
            & $\alpha_\hi$ \quad\quad& $\beta_1$ & $\Re{\alpha_1}$ \quad\quad & $\beta_2$ & $\Re{\alpha_2}$       \\[3pt]
   case 1   & -0.0183378 \quad\quad& 0.3091000 &  -0.2676515 \quad\quad& 0.2971172 & -0.2571380          \\
   case 2   & -0.0133520 \quad\quad& 0.4951000 &  -0.4260816 \quad\quad& 0.1332301 & -0.1099216        \\
   case 3   & -0.0083213 \quad\quad& 0.5741000 &  -0.4905781 \quad\quad& 0.0818596 & -0.0633355         \\ 
   case 4   & -0.0035869 \quad\quad& 0.6311000 &  -0.5359897 \quad\quad& 0.0494499 & -0.0351862         \\ 
   case 5   &  0.0000000 \quad\quad& 0.6683642 &  -0.5651308 \quad\quad& 0.0240432 & -0.0153522
  \end{tabular} 
  \end{subtable}
  \caption{Wavenumbers of stationary vortices for $m=0.5$, $Re=338$ and $2000$.}\label{tab:eigenmodes}
\end{table}

Before presenting the growth-rate correction coefficients and amplitude ratios, a parametric study is performed to monitor the eigenmodes and roughness modes including their wavenumbers and shape functions, as the information is helpful for understanding the sensitive role of roughness. For brevity, the results are presented for $m=0.5$; those for other values of $m$ are similar. 

In order to describe the high-Reynolds-number characteristics of eigenmodes and roughness modes, the concept of a critical level for the stationary crossflow eigenmode and roughness mode is introduced. For the former, the O-S equation (\ref{eq:os}) in the high-Reynolds-number limit reduces to the Rayleigh equation
\begin{equation}
(D^2-\alpha^2-\beta^2)\hat{v}=\frac{\alpha U''+\beta W''}{\alpha U+\beta W}\hat{v}.
\end{equation} 
The singularity of this equation occurs at the critical level $y_c$, which coincides with an inflection point, and thus
\begin{equation}\label{eq:y_c}
\alpha U(y_c)+\beta W(y_c)=0, \quad \alpha U''(y_c)+\beta W''(y_c)=0.
\end{equation}
By eliminating the wavenumbers, we obtain the equation
\begin{equation}
\frac{U(y_c)}{U''(y_c)}=\frac{W(y_c)}{W''(y_c)},
\end{equation}
which allows us to compute $y_c$ from the base flow before solving the eigenvalue problem. In the case of $\phi_0=45^\circ$ and $m=0.5$, it is found that $y_c=1.9915$. 

A roughness mode satisfies the forced Rayleigh equation in the inviscid limit, and it also has a critical level $y_{cw}$, which is given by
\begin{equation}
\alpha_w U(y_{cw})+\beta_w W(y_{cw})=0.
\end{equation}

In figure \ref{fig:sv_roughness_wavenumber}, the roughness wavenumbers $(\beta_{w}, \alpha_{w})$ satisfying the resonance conditions of the triad sum, triad difference and Bragg scattering for stationary vortices are compared with $(\beta_{s}^r, \alpha_{s}^r)$, where $\beta_{s}^r$ and $\alpha_{s}^r$ are the spanwise and chordwise wavenumbers of the right-branch neutral stationary eigenmode respectively. We note that $(\beta_{w}, \alpha_{w})$ in the triad sum are close to $(\beta_{s}^r, \alpha_{s}^r)$ for all growth rates; this closeness occurs also in Bragg scattering around the largest growth rate, and in the triad difference near the smallest growth rate ($-\alpha_\hi=0$). These features of roughness wavenumbers come from the linear instability characteristics of stationary vortices shown in figure \ref{fig:sv_resonance_mechanism}: the wavenumbers of the left-branch neutral eigenmode are much smaller than those of the right-branch neutral eigenmode, and $\alpha_{\hr}$ varies with $\beta$ almost linearly. Therefore, the sum and difference of their wavenumbers are close to $(\beta_{s}^r, \alpha_{s}^r)$ when $\alpha_\hi=0$. The sum of the spanwise wavenumbers of unstable eigenmodes 1 and 2 with equal growth rate turns out to be $\beta^r_s$ approximately, i.e. $\beta_1+\beta_2\approx\beta_s^r$, for all $-\alpha_\hi$. The linear relation implies that the second resonant condition, $\alpha_1+\alpha_2=\alpha_s^r$, is automatically satisfied. Since the triad sum degenerates to Bragg scattering at the largest growth rate, where eigenmodes 1 and 2 are identical, the roughness wavenumbers in Bragg scattering around the largest growth rate are close to $(\beta_{s}^r, \alpha_{s}^r)$. As the growth rate decreases, the roughness wavenumbers in Bragg scattering deviate from $(\beta_{s}^r, \alpha_{s}^r)$, whereas the opposite is true for the triad difference. With the Reynolds number increased, the roughness wavenumbers in the triad sum become closer to $(\beta_{s}^r, \alpha_{s}^r)$ generally, as indicated by figure \ref{fig:sv_roughness_wavenumber}.

\begin{figure}
% \psfrag{y}{$y$}
% \psfrag{u}[][]{$|\hat{u}|$}
% \psfrag{v}[][]{$|\hat{v}|$}
% \psfrag{w}[][]{$|\hat{w}|$}
% \psfrag{A}[][]{$(a)$}
% \psfrag{B}[][]{$(b)$}
% \psfrag{P}[][]{$(1)$}
% \psfrag{Q}[][]{$(2)$}
% \psfrag{l}{$y_c$}
\centerline{\includegraphics{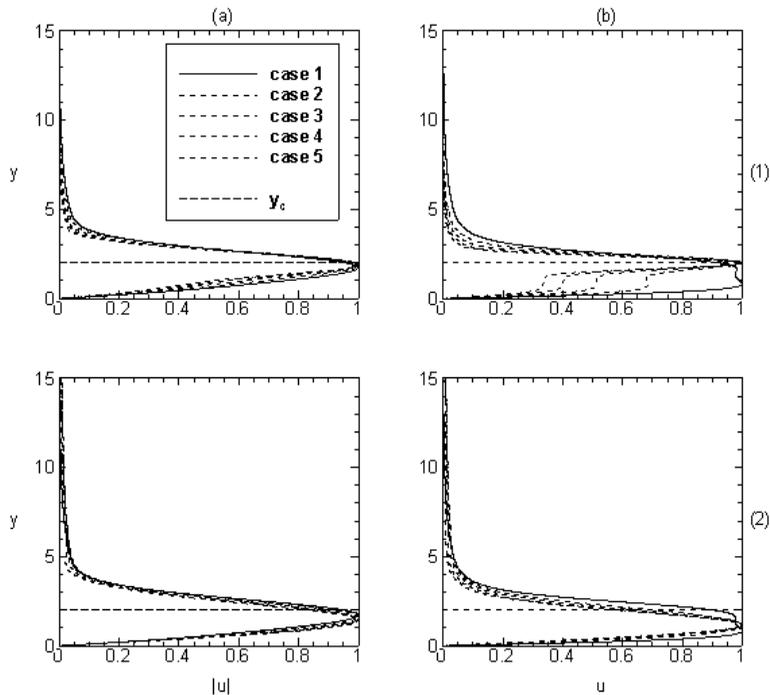}}
\caption{The eigenfunctions of the eigenmodes on the $(1)$ right and $(2)$ left branches: $(a)$ $Re=338$; $(b)$ $Re=2000$.}
\label{fig:sv_eigenfunction_eigenmodes}
\end{figure}

The shape functions of the stationary vortices and roughness modes in a resonant triad are presented for the five cases marked by the thin vertical lines in figure \ref{fig:sv_roughness_wavenumber}. The growth rates and wavenumbers of these eigenmodes are listed in table \ref{tab:eigenmodes}. Among these cases, the stationary vortices are nearly most unstable in case 1 and neutral in case 5. The exactly most unstable stationary eigenmode is not chosen because the corresponding difference of its wavenumbers is zero, which prohibits use of the forced O-S equation for computing the roughness-induced perturbation.  The eigenfunctions of the eigenmodes 1 located on the right branch are shown in figure \ref{fig:sv_eigenfunction_eigenmodes}(1) for $Re=338$ and $Re=2000$. Note that eigenfunctions are normalized by the maximum value of $|\hat{u}|$. For brevity, only the chordwise disturbance velocity $|\hat{u}|$ is displayed since the spanwise disturbance velocity $|\hat{w}|$ closely resembles $|\hat{u}|$. At $Re=338$, the vertical position $y_{m}$ of the maximum value of the eigenfunction for the neutral eigenmode of case 5 is closest  to the critical level $y_c$ among all cases. Although the eigenmodes of cases 1-4 are unstable, their $y_m$ are also fairly close to $y_c$. As the Reynolds number increases, the eigenfunctions in cases 2-5 display a two-layer structure. Inspecting, for example, the eigenfunctions at $Re=2000$ of eigenmodes with low growth rates, one observes a wall layer, where viscosity cannot be ignored. Towards the outer edge of the wall layer, the eigenfunctions tend to constants. The perturbation concentrates in the so-called critical layer surrounding $y_c$. The positions of the maxima of the eigenfunctions are extremely close to $y_c$. For the nearly most unstable eigenmode in case 1, the wall layer and the critical layer are not distinct, which is expected since the critical layer concept is not attainable. The above observations suggest that right-branch stationary eigenmodes are essentially inviscid. The critical-layer concept, which originates from inviscid instability for neutral stationary modes, is applicable to right-branch unstable stationary eigenmodes provided that their growth rates are not too large, and the Reynolds number is high enough. The eigenfunctions of the eigenmodes 2 located on the left branch for the five cases are shown in figure \ref{fig:sv_eigenfunction_eigenmodes}(2). Unlike eigenmodes 1, the eigenfunctions of these modes do not exhibit a two-layer structure. Nevertheless, the positions of their maxima are fairly close to $y_c$ as well.

\begin{figure}
% \psfrag{y}{$y$}
% \psfrag{u}[][]{$|\hat{u}_w|$}
% \psfrag{v}[][]{$|\hat{v}_w|$}
% \psfrag{w}[][]{$|\hat{w}_w|$}
% \psfrag{A}[][]{$(a)$}
% \psfrag{B}[][]{$(b)$}
% \psfrag{l}{$y_c$}
% \psfrag{p}{$y_{cw1}$}
% \psfrag{q}{$y_{cw3}$}
% \psfrag{t}{$y_{cw5}$}
\centerline{\includegraphics{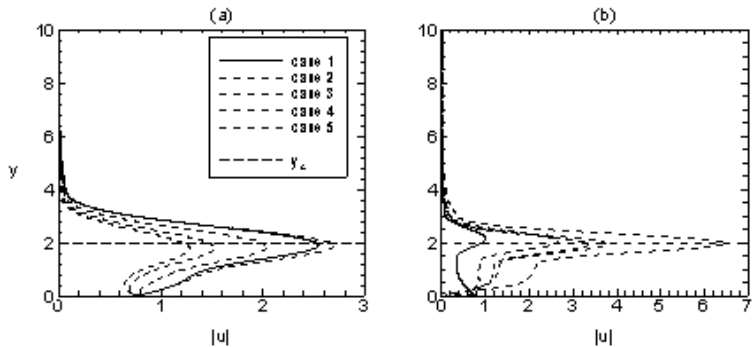}}
\caption{Distribution of the roughness modes in the triad sum (1+2) interaction: $(a)$ $Re=338$; $(b)$ $Re=2000$.}
\label{fig:sv_shape_roughness_1+2}
\end{figure}

\begin{figure}
% \psfrag{y}{$y$}
% \psfrag{p}[][]{$|\hat{u}|$}
% \psfrag{q}[][]{$|\hat{v}|$}
% \psfrag{t}[][]{$|\hat{w}|$}
\centerline{\includegraphics{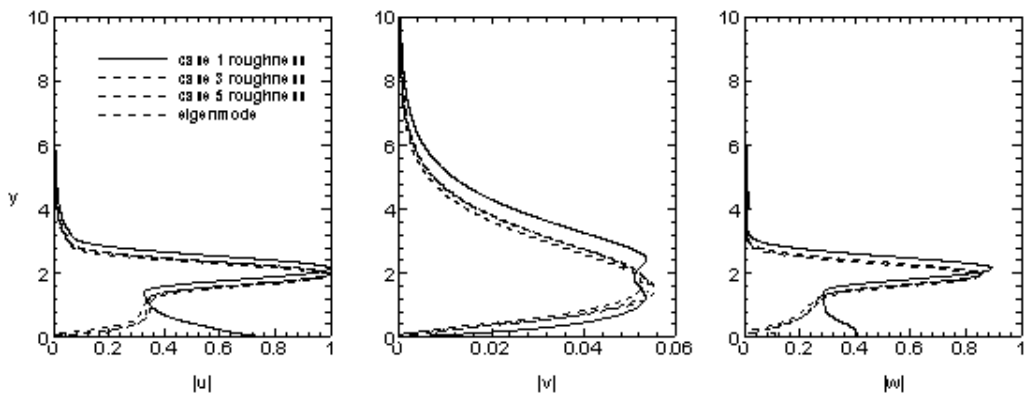}}
\caption{Comparison of the roughness modes in the triad sum of cases 1, 3, and 5 at $Re=2000$ with the eigenfunction of the right-branch neutral eigenmode.}
\label{fig:sv_eigenmode_roughness}
\end{figure}

The shape functions of the roughness modes for the chosen cases are now presented. Figure \ref{fig:sv_shape_roughness_1+2} shows the results for the roughness modes interacting with stationary vortices through the triad sum. For all the cases, the roughness modes share the same boundary value at the wall surface, whose magnitude is of $\textit{O}(10^{-1})$. The difference of the shape functions is due to the roughness wavenumbers. Relative to the size at the wall surface, the peak values of the roughness modes are amplified significantly in the critical layer. The reason for this is that the corresponding roughness wavenumbers in the triad sum are very close to $(\beta_s^r, \alpha_s^r)$ as shown in figure \ref{fig:sv_roughness_wavenumber}, that is, the roughness modes are in near-resonance with the right-branch neutral stationary eigenmode. At $Re=338$, as the growth rate increases from case 5 to case 2, and the roughness wavenumbers $(\beta_w, \alpha_w)$ in the triad sum become closer to $(\beta_s^r, \alpha_s^r)$ (figure \ref{fig:sv_roughness_wavenumber}($a$)), the peak values of the corresponding shape functions become larger and their vertical positions move closer to $y_c$ as displayed in figure \ref{fig:sv_shape_roughness_1+2}($a$). At $Re=2000$, a similar trend can be observed from case 1 to case 3 (figure \ref{fig:sv_shape_roughness_1+2}($b$)) with $(\beta_w, \alpha_w)$ getting closer to $(\beta_s^r, \alpha_s^r)$ as indicated by figure \ref{fig:sv_roughness_wavenumber}($b$). At higher Reynolds numbers, the peak values of the roughness modes are greater in general, because the corresponding roughness wavenumbers are closer to $(\beta_s^r, \alpha_s^r)$; see figure \ref{fig:sv_roughness_wavenumber}($b$). 

\begin{figure}
% \psfrag{y}{$y$}
% \psfrag{u}[][]{$|\hat{u}_w|$}
% \psfrag{v}[][]{$|\hat{v}_w|$}
% \psfrag{w}[][]{$|\hat{w}_w|$}
% \psfrag{A}[][]{$(a)$}
% \psfrag{B}[][]{$(b)$}
% \psfrag{P}[][]{$(1)$}
% \psfrag{Q}[][]{$(2)$}
% \psfrag{l}{$y_c$}
% \psfrag{p}{$y_{cw1}$}
% \psfrag{q}{$y_{cw3}$}
% \psfrag{t}{$y_{cw5}$}
\centerline{\includegraphics{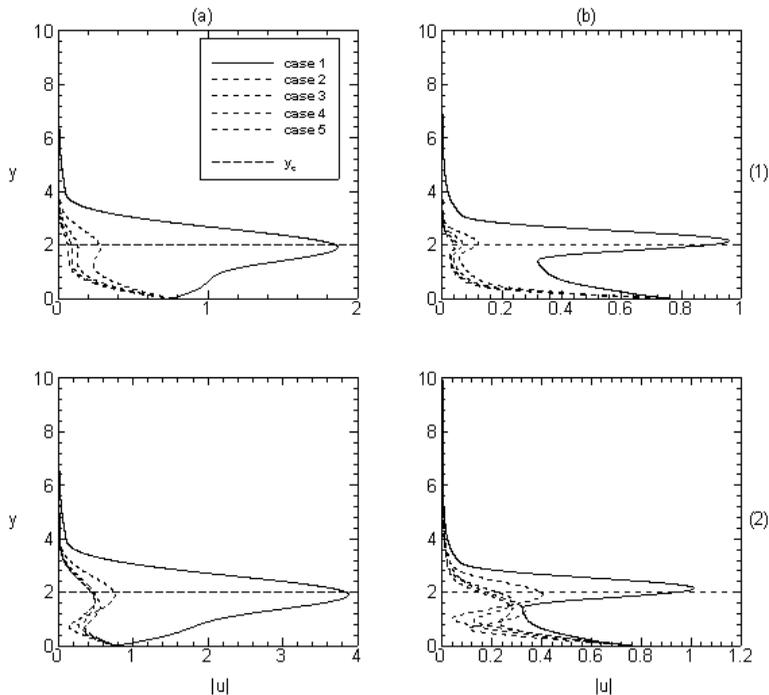}}
\caption{Distribution of the roughness modes in the Bragg scattering for the $(1)$ right-branch and $(2)$ left-branch eigenmodes: $(a)$ $Re=338$; $(b)$ $Re=2000$.}
\label{fig:sv_shape_roughness_bragg-scattering}
\end{figure}

\begin{figure}
% \psfrag{y}{$y$}
% \psfrag{u}[][]{$|\hat{u}_w|$}
% \psfrag{v}[][]{$|\hat{v}_w|$}
% \psfrag{w}[][]{$|\hat{w}_w|$}
% \psfrag{A}[][]{$(a)$}
% \psfrag{B}[][]{$(b)$}
% \psfrag{l}{$y_c$}
% \psfrag{p}{$y_{cw1}$}
% \psfrag{q}{$y_{cw3}$}
% \psfrag{t}{$y_{cw5}$}
\centerline{\includegraphics{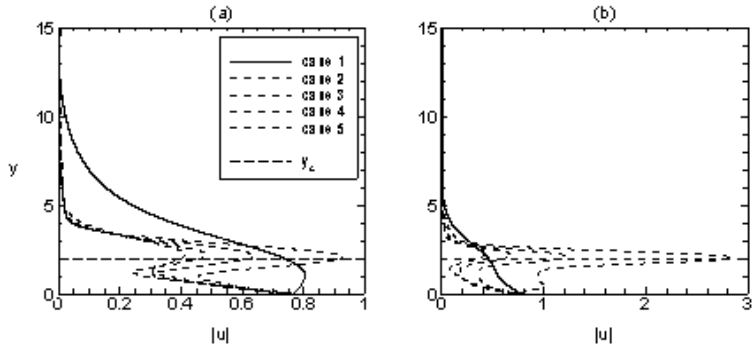}}
\caption{Distribution of the roughness modes in the triad difference (1-2) interaction: $(a)$ $Re=338$; $(b)$ $Re=2000$.}
\label{fig:sv_shape_roughness_1-2}
\end{figure}

In figure \ref{fig:sv_eigenmode_roughness}, the shape functions of the roughness modes in the triad sum of cases 1, 3 and 5 at $Re=2000$ are compared with those of the right-branch neutral stationary  eigenmode. Note that the shape functions of the roughness modes are normalized by the maximum value of $|\hat{u}_w|$. The transverse distributions of the roughness modes almost overlap with the eigenfunction of the right-branch neutral stationary  crossflow vortices. For case 1, where the growth rate is almost the largest, the resemblance between the shape functions is not so remarkable as that for lower growth rates, but in the critical layer the roughness mode is still similar to the right-branch neutral stationary eigenmode.

\begin{figure}
% \psfrag{y}{$y$}
% \psfrag{u}[][]{$|f_{2-w}^x|$}
% \psfrag{v}[][]{$|f_{2-w}^y|$}
% \psfrag{w}[][]{$|f_{2-w}^z|$}
% \psfrag{A}{$(a)$}
% \psfrag{B}{$(b)$}
% \psfrag{C}{$(c)$}
\centerline{\includegraphics[scale=0.85]{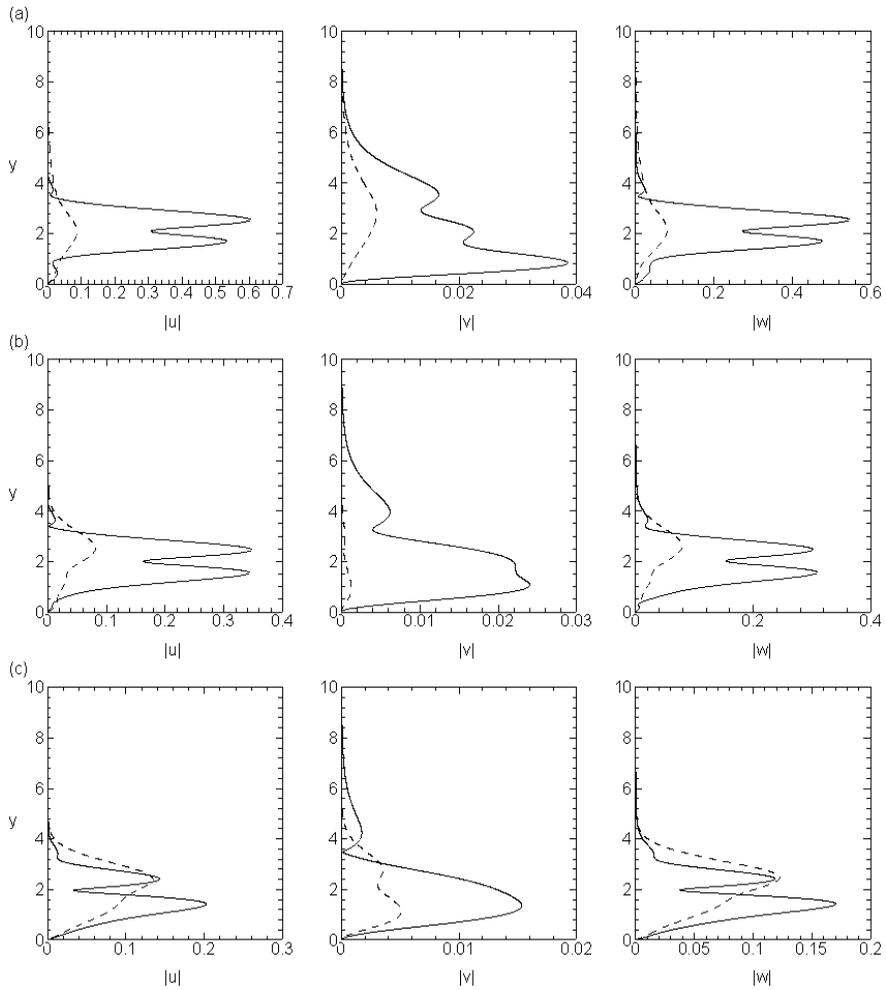}}
\caption{Distribution of the Reynolds stresses induced by eigenmode 2 interacting with the roughness modes at $Re=338$: solid lines, roughness 1+2; dashed lines, roughness 1-2. $(a)$ case 1; $(b)$ case 3; $(c)$ case 5.}
\label{fig:sv_re=338_reynolds_stress_1}
\end{figure}

\begin{figure}
% \psfrag{y}{$y$}
% \psfrag{u}[][]{$|f_{1-w}^x|$}
% \psfrag{v}[][]{$|f_{1-w}^y|$}
% \psfrag{w}[][]{$|f_{1-w}^z|$}
% \psfrag{A}{$(a)$}
% \psfrag{B}{$(b)$}
% \psfrag{C}{$(c)$}
\centerline{\includegraphics[scale=0.85]{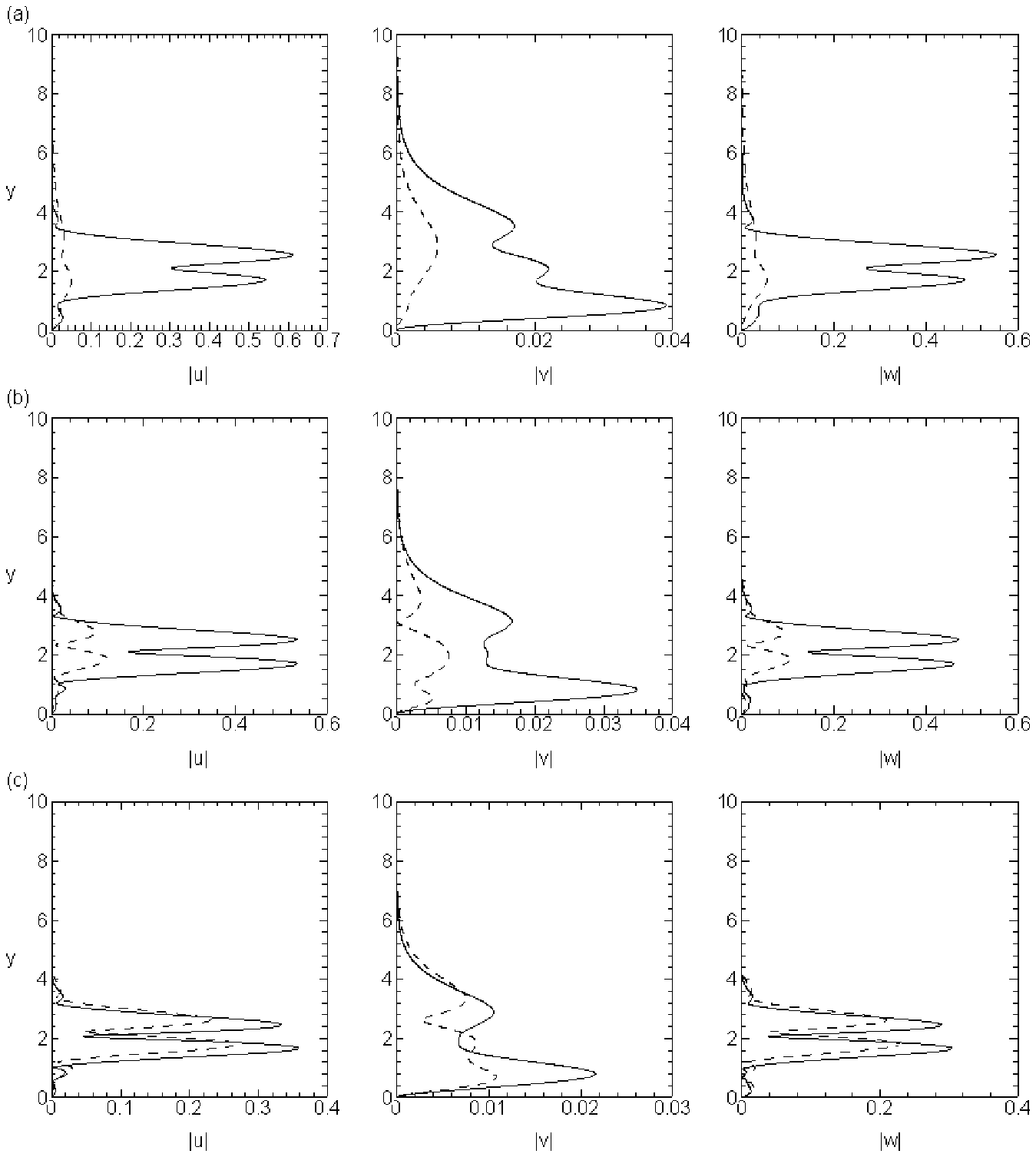}}
\caption{Distribution of the Reynolds stresses induced by eigenmode 1 interacting with the roughness modes at $Re=338$: solid lines, roughness 1+2; dashed lines, roughness 1-2. $(a)$ case 1; $(b)$ case 3; $(c)$ case 5.}
\label{fig:sv_re=338_reynolds_stress_2}
\end{figure}

\begin{figure}
% \psfrag{y}{$y$}
% \psfrag{u}[][]{$|f_{2-w}^x|$}
% \psfrag{v}[][]{$|f_{2-w}^y|$}
% \psfrag{w}[][]{$|f_{2-w}^z|$}
% \psfrag{A}{$(a)$}
% \psfrag{B}{$(b)$}
% \psfrag{C}{$(c)$}
\centerline{\includegraphics[scale=0.85]{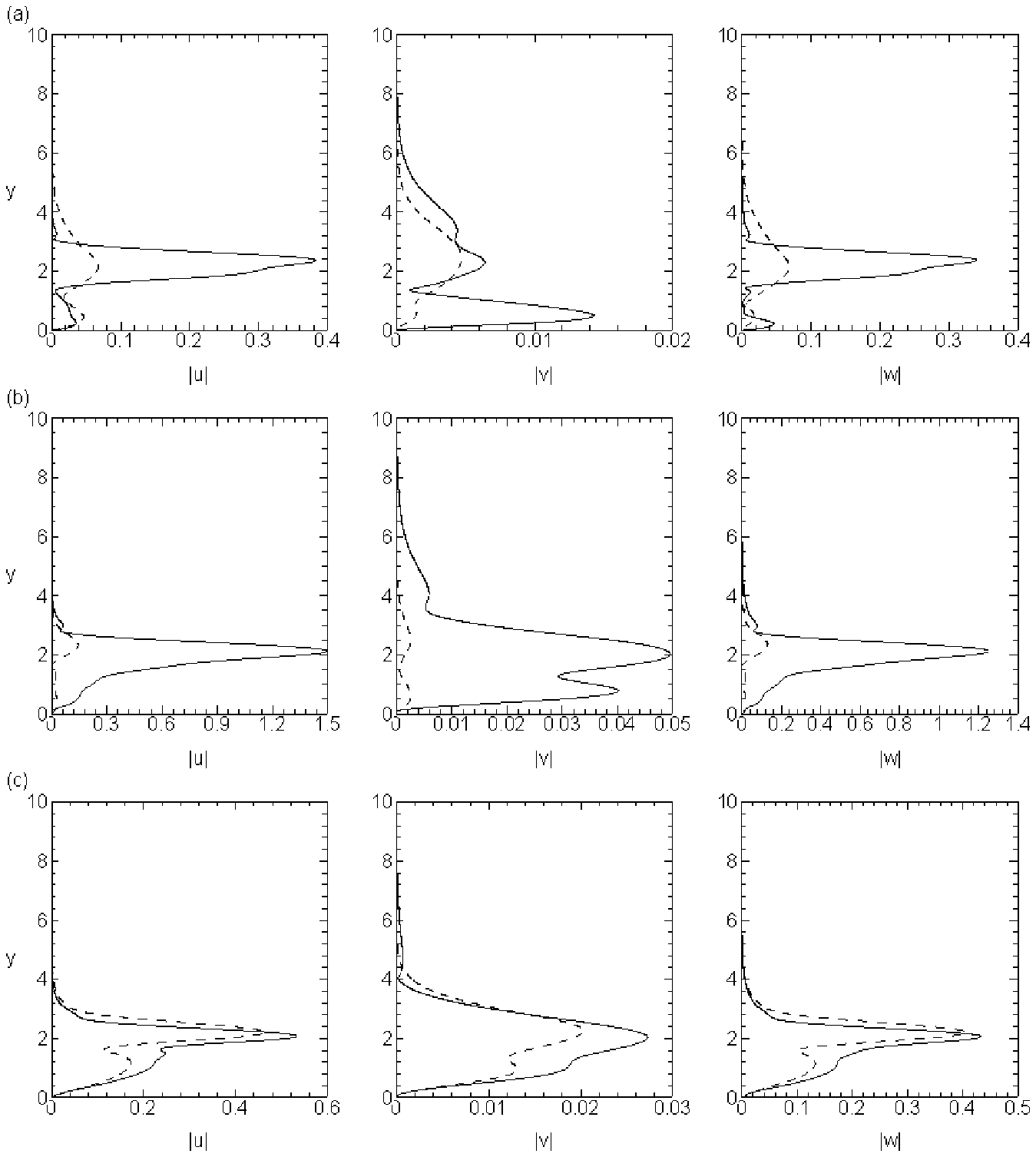}}
\caption{Distribution of the Reynolds stresses induced by eigenmode 2 interacting with the roughness modes at $Re=2000$: solid lines, roughness 1+2; dashed lines, roughness 1-2. $(a)$ case 1; $(b)$ case 3; $(c)$ case 5.}
\label{fig:sv_re=2000_reynolds_stress_1}
\end{figure}

\begin{figure}
% \psfrag{y}{$y$}
% \psfrag{u}[][]{$|f_{1-w}^x|$}
% \psfrag{v}[][]{$|f_{1-w}^y|$}
% \psfrag{w}[][]{$|f_{1-w}^z|$}
% \psfrag{A}{$(a)$}
% \psfrag{B}{$(b)$}
% \psfrag{C}{$(c)$}
\centerline{\includegraphics[scale=0.85]{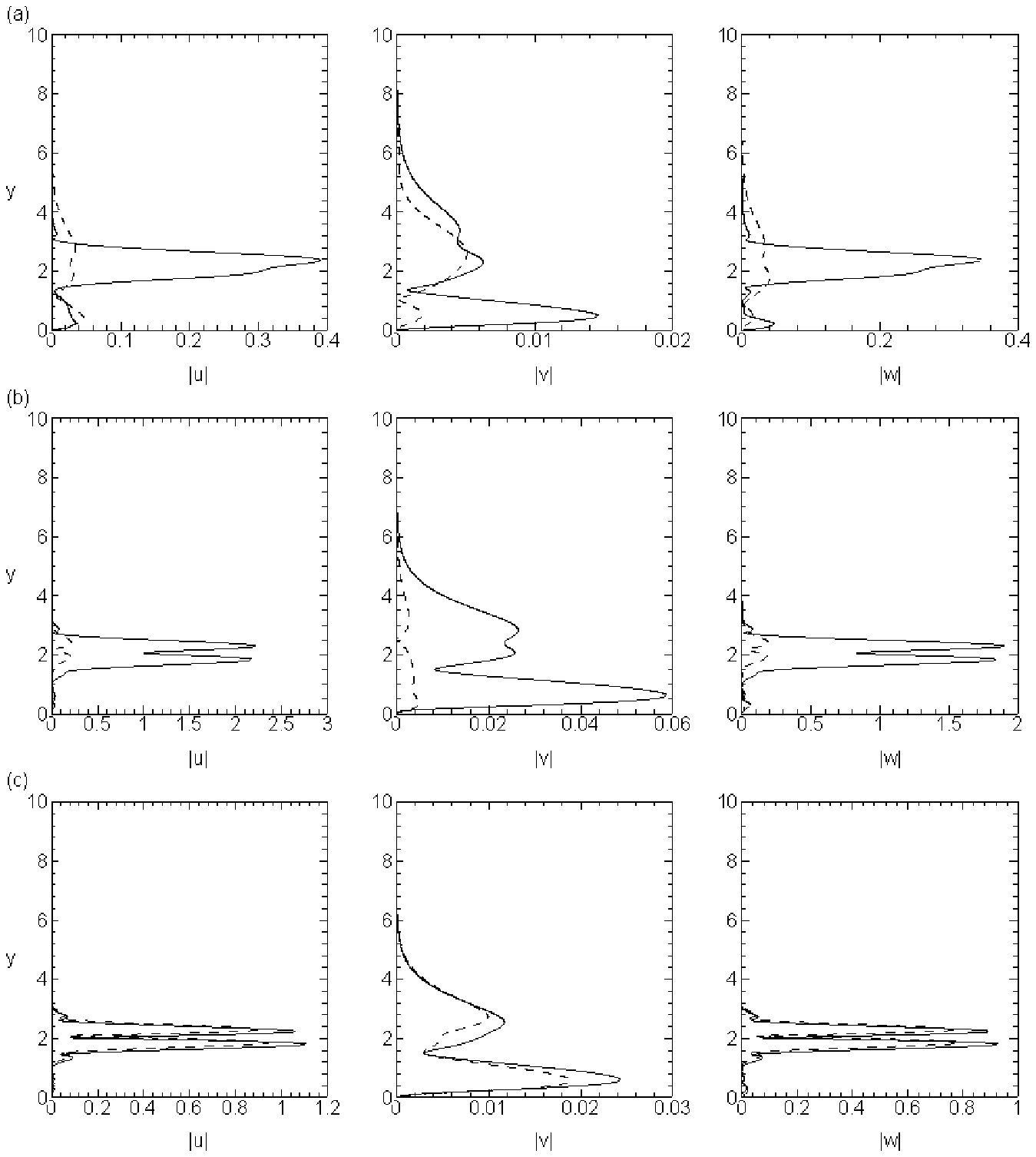}}
\caption{Distribution of the Reynolds stresses induced by eigenmode 1 interacting with the roughness modes at $Re=2000$: solid lines, roughness 1+2; dashed lines, roughness 1-2. $(a)$ case 1; $(b)$ case 3; $(c)$ case 5.}
\label{fig:sv_re=2000_reynolds_stress_2}
\end{figure}

Figure \ref{fig:sv_shape_roughness_bragg-scattering} shows the roughness modes in Bragg scattering for the right-branch and left-branch stationary eigenmodes. Among the five chosen cases, only the roughness mode of case 1 exhibits a critical-layer amplification for both the right-branch and left-branch eigenmodes. The reason is that only in this case are the wavenumbers $(\beta_w, \alpha_w)$ of the roughness mode sufficiently close to $(\beta_s^r, \alpha_s^r)$; see figure \ref{fig:sv_roughness_wavenumber}. At $Re=2000$, the roughness mode of case 1 in Bragg scattering displays a two-layer structure as those in the triad sum. As the growth rate decreases, the roughness wavenumbers deviate from $(\beta_s^r, \alpha_s^r)$ (figure \ref{fig:sv_roughness_wavenumber}), which means that the corresponding response is non-resonant. There is no critical-layer amplification for cases 2-5, but there still exists a local peak near the critical level. By comparing the maximum values of the roughness modes of case 1 in Bragg scattering at $Re=338$ and $2000$, one finds that the maximum values at a lower Reynolds number are much larger than those at a higher Reynolds number especially for the left-branch eigenmode. Bragg scattering is therefore expected to have a greater effect on the nearly most unstable stationary vortices at lower Reynolds numbers. This will be demonstrated directly after computing the growth-rate corrections.

Figure \ref{fig:sv_shape_roughness_1-2} shows the shape functions of the roughness modes in the triad difference. Just as those in Bragg scattering, the roughness modes in the triad difference fall into two categories: near-resonant and non-resonant. The roughness modes of case 5 have large critical-layer amplification, and are therefore near-resonant; this is expected since their wavenumbers are very close to $(\beta_s^r, \alpha_s^r)$, as shown in figure \ref{fig:sv_roughness_wavenumber}. As the Reynolds number increases, the wavenumbers of roughness 1-2 in case 5 become closer to $(\beta_s^r, \alpha_s^r)$, leading to greater critical-layer amplification. The roughness modes in cases 1-4 are non-resonant since their wavenumbers differ appreciably from $(\beta_s^r, \alpha_s^r)$, and their shape functions do not show amplification in the critical layer. 

The main difference between the triad sum and difference for the same pair of stationary vortices is the roughness mode involved. The Reynolds stresses induced by the interaction between the eigenmode and the roughness mode can be very different in these two mechanisms. Figures \ref{fig:sv_re=338_reynolds_stress_1} and \ref{fig:sv_re=338_reynolds_stress_2} present the distribution of the Reynolds stresses induced by the eigenmodes 2 and 1 interacting with the roughness mode respectively at $Re=338$. For cases 1 and 3, the Reynolds stresses in the triad sum are significantly larger than those in the triad difference. A similar feature is also observed at a higher Reynolds number $Re=2000$; see figures \ref{fig:sv_re=2000_reynolds_stress_1} and \ref{fig:sv_re=2000_reynolds_stress_2}.
The huge disparity in Reynolds stresses is due to the fact that the corresponding roughness modes of cases 1 and 3 in the triad sum are near-resonant, while those in the triad difference are non-resonant. For case 5, where neutral eigenmodes are involved, the roughness modes for the triad sum and difference are both near-resonant, and so the Reynolds stresses are comparable.

\begin{figure}
% \psfrag{x}[][]{$-\alpha_\hi$}
% \psfrag{y}{$\lambda_\hr$}
% \psfrag{A}[][]{$(a)$}
% \psfrag{B}[][]{$(b)$}
% \psfrag{o}{$(1)$}
% \psfrag{p}{$(2)$}
% \psfrag{q}{$(3)$}
\centerline{\includegraphics[scale=0.9]{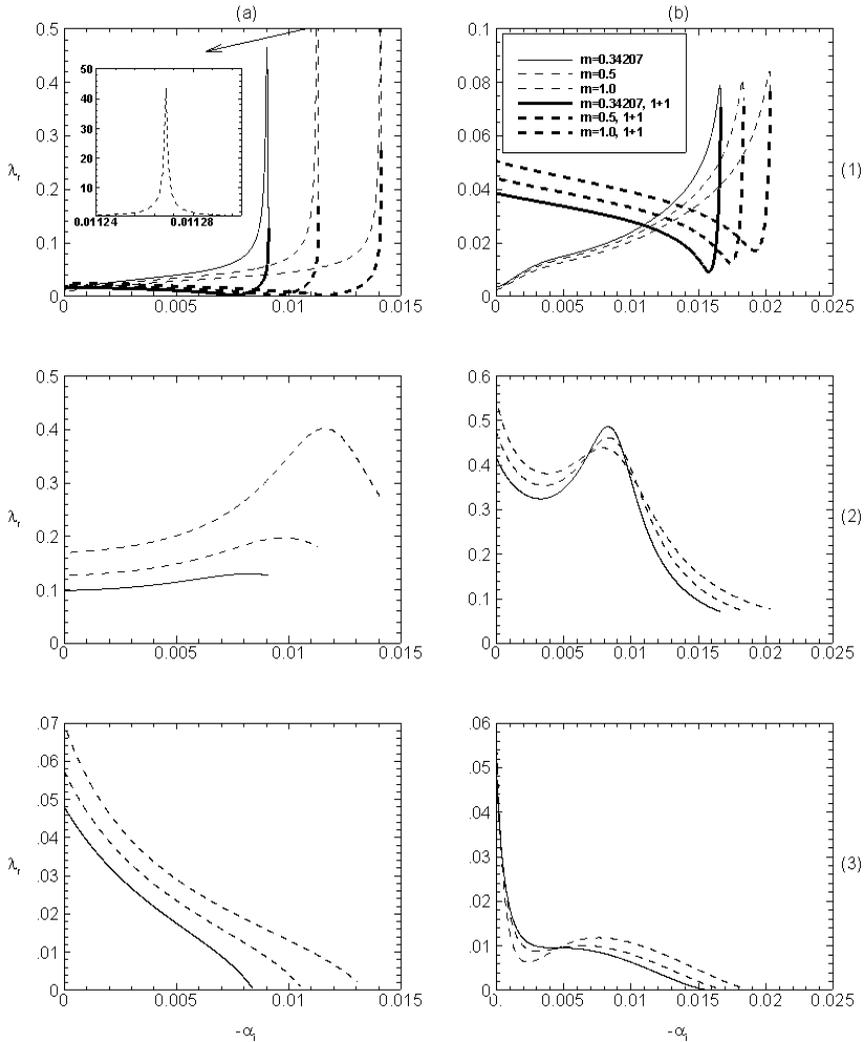}}
\caption{Variation of $\lambda_\hr$ with $-\alpha_\hi$ for stationary vortices interacting with single-wavenumber roughness through $(1)$ Bragg scattering, $(2)$ triad sum and $(3)$ triad difference. $(a)$ $Re=338$; $(b)$ $Re=2000$.}
\label{fig:sv_lambda}
\end{figure}

Figure \ref{fig:sv_lambda} shows the variation of the growth-rate correction coefficient $\lambda_\hr$ with the leading-order growth rate ($-\alpha_\hi$) for stationary vortices interacting with single-wavenumber roughness through Bragg scattering, triad sum and triad difference for the three acceleration parameters and two Reynolds numbers. For Bragg scattering (figure \ref{fig:sv_lambda}($1$)), the curve representing the correction coefficient has two branches, because the left and right-branch modes are affected independently. The two branches merge at the maximum growth rate. The correction coefficients are larger for all left-branch modes at $Re=338$ (figure \ref{fig:sv_lambda}($1a$)), but at $Re=2000$, larger $\lambda_\hr$ is observed for the left-branch eigenmodes with relatively large $-\alpha_\hi$, and this switches to the right-branch modes when $-\alpha_\hi$ is small (figure \ref{fig:sv_lambda}($1b$)). The correction coefficient around the maximum growth rate is much larger than those at smaller growth rates because the roughness modes involved are near-resonant. This means that the most or nearly most unstable stationary vortices can become even more unstable due to Bragg scattering. At $Re=338$ the peak value of $\lambda_\hr$ through Bragg scattering is extremely large (about $45$, much greater than that at $Re=2000$). This is attributed to the fact that the corresponding roughness wavenumbers are almost equal to $(\beta_s^r, \alpha_s^r)$, and so the forcing is almost in exact resonance with the right-branch neutral stationary  eigenmode. For suitable base-flow parameters and stationary eigenmode, an exact resonance may occur and Bragg scattering leads to an infinitely large $\lambda_\hr$, in which case the problem has to be regularized by introducing appropriate physical effects, which can be the non-parallelism of the base flow. 

\begin{figure}
% \psfrag{x}[][]{$-\alpha_\hi$}
% \psfrag{y}{$\lambda_\hr$}
% \psfrag{A}[][]{$(a)$}
% \psfrag{B}[][]{$(b)$}
% \psfrag{r}{$(1)$}
% \psfrag{s}{$(2)$}
% \psfrag{w}{$(3)$}
\centerline{\includegraphics[scale=0.9]{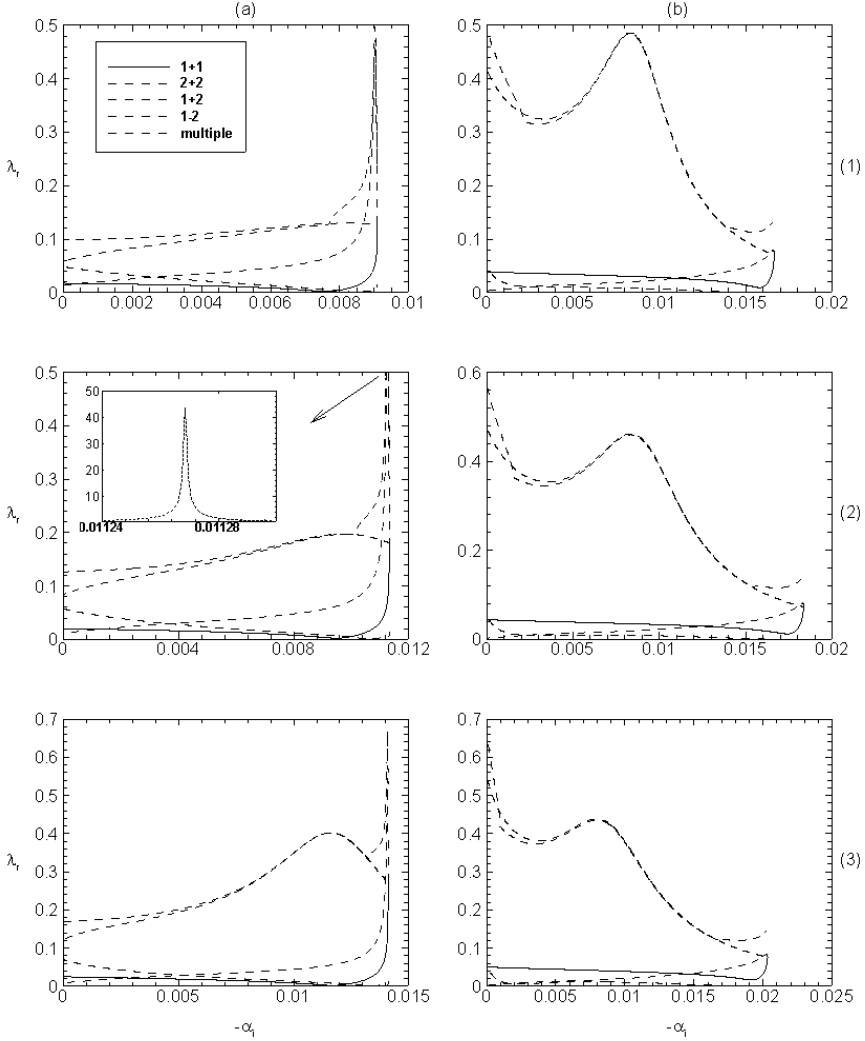}}
\caption{Comparisons of $\lambda_\hr$ for stationary vortices interacting with multiple-wavenumber and single-wavenumber roughness: $(a)$ $Re=338$; $(b)$ $Re=2000$; $(1)$ $m=0.34207$; $(2)$ $m=0.5$; $(3)$ $m=1.0$.}
\label{fig:sv_multiple_single_lambda}
\end{figure}

\begin{figure}
% \psfrag{x}[][]{$-\alpha_\hi$}
% \psfrag{y}{$\rho$}
% \psfrag{A}[][]{$(a)$}
% \psfrag{B}[][]{$(b)$}
% \psfrag{o}{$(1)$}
% \psfrag{p}{$(2)$}
% \psfrag{q}{$(3)$}
\centerline{\includegraphics[scale=0.9]{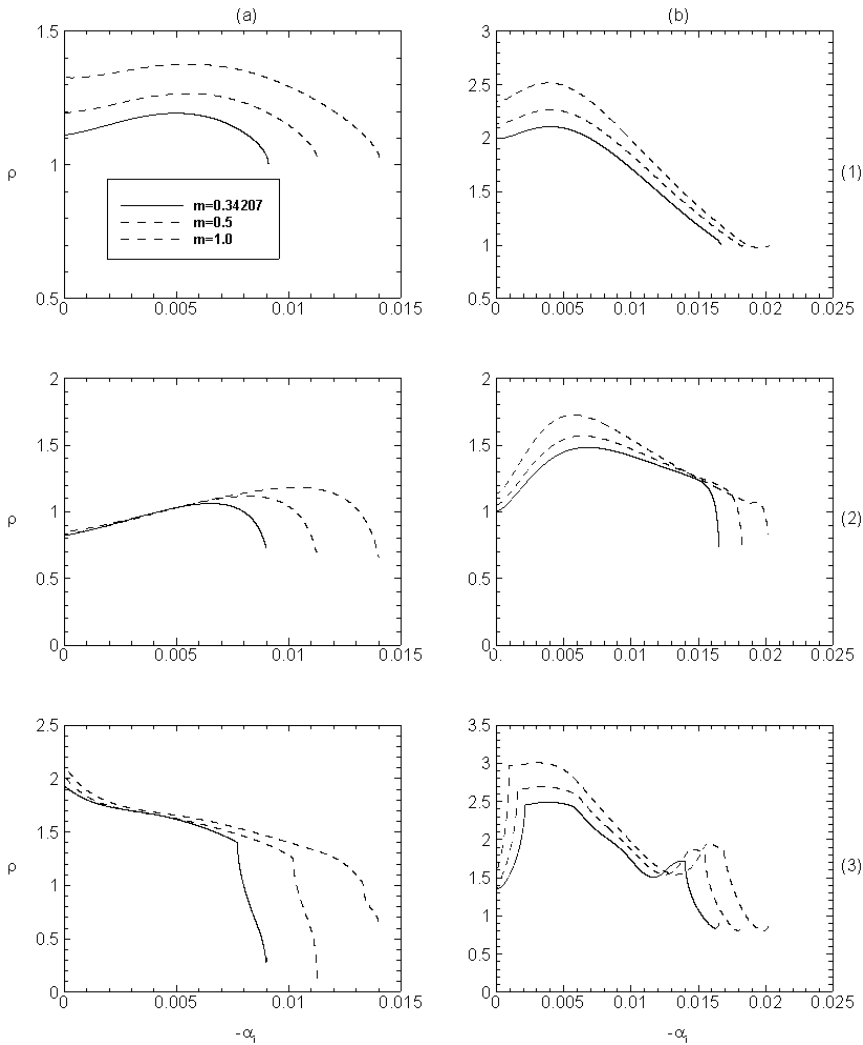}}
\caption{Variation of $\rho$ with $-\alpha_\hi$ for stationary vortices interacting with single-wavenumber roughness through $(1)$ triad sum, $(2)$ triad difference and with $(3)$ multiple-wavenumber roughness. $(a)$ $Re=338$; $(b)$ $Re=2000$.}
\label{fig:sv_amplitude_ratio}
\end{figure}

The correction coefficient due to the triad sum is shown in figure \ref{fig:sv_lambda}$(2)$. At $Re=338$, the correction coefficients for higher growth rates are generally larger than those for lower growth rates, whereas at $Re=2000$ the opposite is true. The correction is rather substantial for the majority of ($-\alpha_\hi$) values, which is because the roughness wavenumbers are close to $(\beta_s^r, \alpha_s^r)$. As $m$ increases, $\lambda_\hr$ becomes larger. Since the case with $m=1$ is representative of the flow condition near the leading edge, the above trend is consistent with the experimental observation that roughness in that region plays a more sensitive role. Figure \ref{fig:sv_lambda}$(3)$ displays the correction coefficient due to the triad difference. The correction coefficient attains its maximum at $-\alpha_\hi=0$ since the roughness mode is in near-resonance with the right-branch neutral stationary crossflow vortices. For the majority of ($-\alpha_\hi$) values, the correction coefficient due to the triad difference is merely of $\textit{O}(10^{-2})$, while that due to the triad sum is of $\textit{O}(10^{-1})$; the reason for this remarkable disparity was given above. As we will show later, the triad sum can be a very efficient mechanism, through which micron-sized roughness creates a more-or-less $\textit{O}(1)$ correction to the leading-order growth rate of stationary vortices.

The above results are for single-wavenumber roughness. As we noted earlier, a pair of stationary crossflow modes with equal growth rate can interact with four roughness components. Calculations were carried out for this extended case assuming that all roughness components have the same amplitude. The resulting correction coefficients are compared in figure \ref{fig:sv_multiple_single_lambda} with those for single-wavenumber roughness. A main feature is that the $\lambda_\hr$ for the multiple-wavenumber interaction is similar to that for the single-wavenumber interaction that dominates among the four interactions. At $Re=338$, interactions 2+2 and 1+2 dominate at high and low growth rates respectively; at $Re=2000$, the dominant one is always 1+2. These results indicate the special importance of Bragg scattering for stationary eigenmodes with high growth rates at low Reynolds numbers, and the strong effect of the triad sum on the majority of stationary eigenmodes. For larger $m$, the overall correction coefficient increases.

Besides the correction to the growth rates, distributed roughness determines the amplitude ratio of the eigenmodes involved in the resonance. Figure \ref{fig:sv_amplitude_ratio} shows the amplitude ratio $\rho$ for single-wavenumber and multiple-wavenumber roughnesses. Note that $\rho$ is taken as the ratio of the amplitude of the right-branch eigenmode to that of the left-branch eigenmode. For the triad sum $\rho >1$ (figure \ref{fig:sv_amplitude_ratio}(1)), which means that the right-branch stationary eigenmode is stronger than its left-branch counterpart. As the acceleration parameter $m$ increases, so does $\rho$. For the triad difference (figure \ref{fig:sv_amplitude_ratio}(2)), the amplitude ratio at intermediate growth rates is larger than that at low and high growth rates. The amplitude ratio of stationary vortices interacting with multiple-wavenumber roughness is shown in figure \ref{fig:sv_amplitude_ratio}(3). Interestingly, although the correction coefficient is determined by the dominant interaction, $\rho$ is not; minor interactions such the triad difference appeal to affect the amplitude ratio substantially.

\subsection{Resonant interactions of travelling vortices with roughness modes}

\begin{figure}
% \psfrag{A}[][]{$(a)$}
% \psfrag{B}[][]{$(b)$}
% \psfrag{x}[][]{$-\alpha_\hi$}
% \psfrag{y}{$\beta_w$}
% \psfrag{z}{$\alpha_w$}
\centerline{\includegraphics[scale=1]{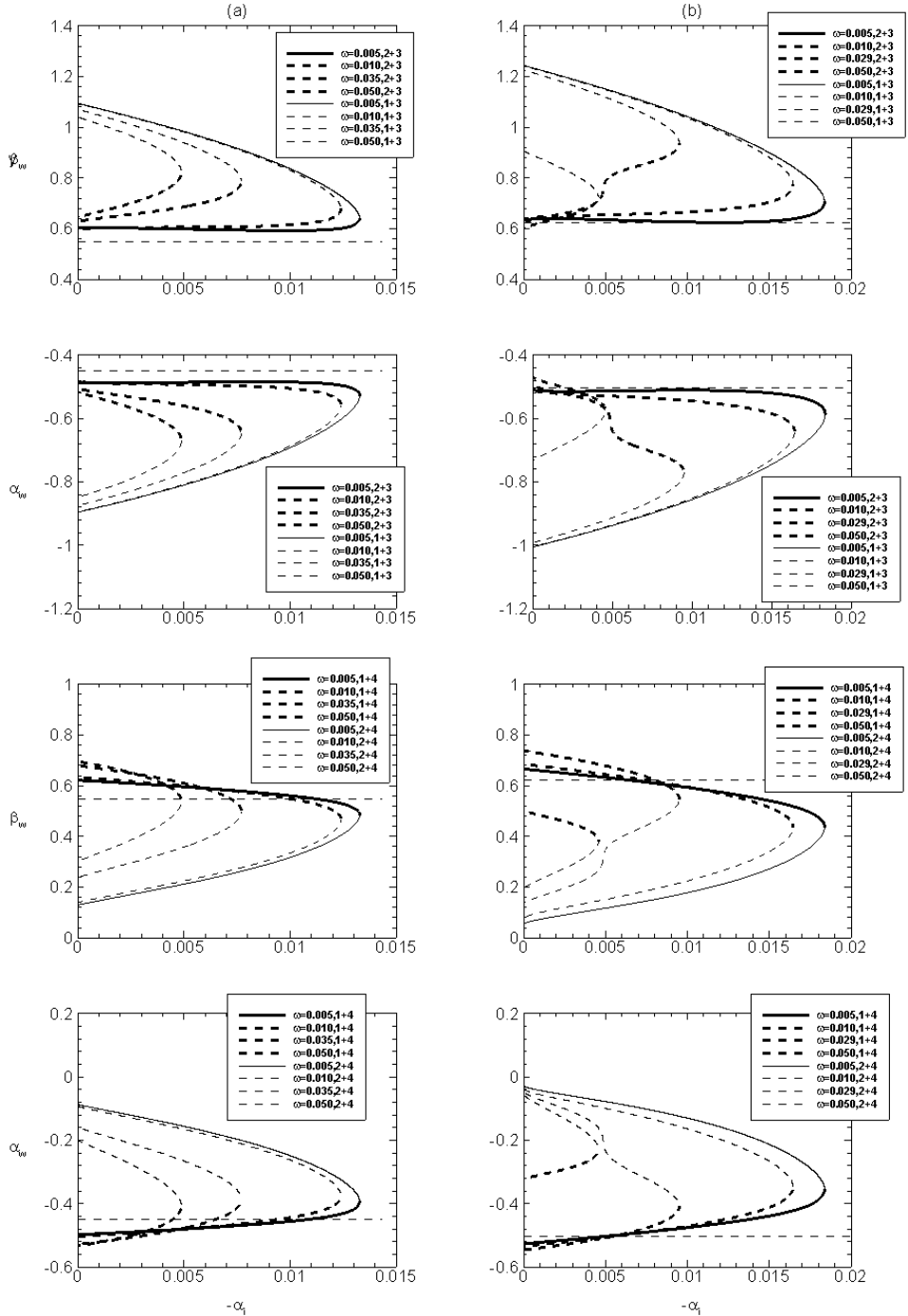}}
\caption{The wavenumbers of the roughness modes in the triad-sum interactions with travelling vortices for $m=1.0$: $(a) Re=338$; $(b) Re=2000$. The long dashed lines stand for $(\beta_s^r, \alpha_s^r)$.}
\label{fig:tw_triad_sum_roughness_wavenumber}
\end{figure}

\begin{figure}
% \psfrag{A}[][]{$(a)$}
% \psfrag{B}[][]{$(b)$}
% \psfrag{x}[][]{$-\alpha_\hi$}
% \psfrag{y}{$\beta_w$}
% \psfrag{z}{$\alpha_w$}
% \psfrag{w}[][]{$\omega$}
\centerline{\includegraphics[scale=1]{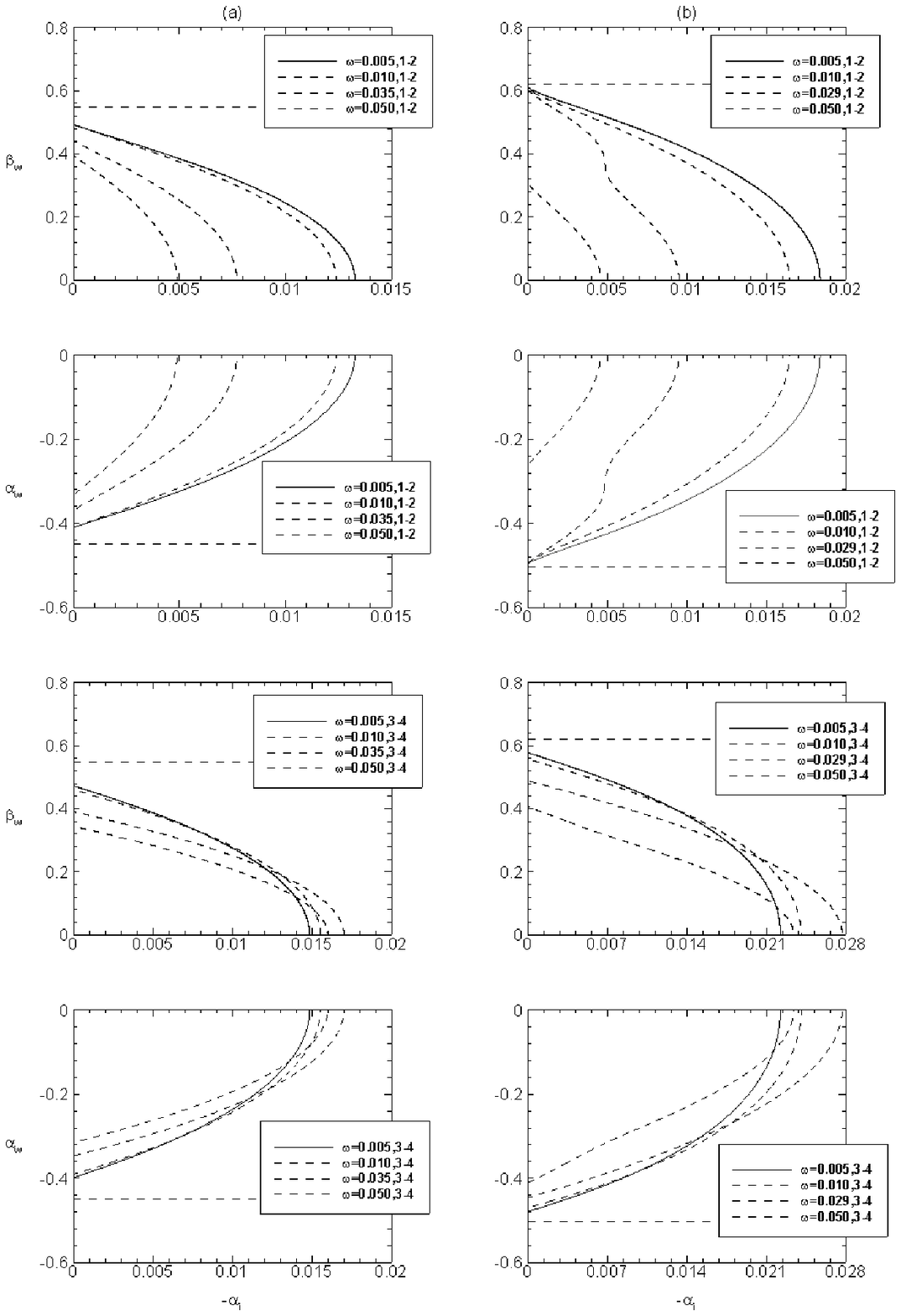}}
\caption{The wavenumbers of the roughness modes in the triad-difference interactions with travelling vortices for $m=1.0$: $(a) Re=338$; $(b) Re=2000$. The long dashed lines stand for $(\beta_s^r, \alpha_s^r)$.}
\label{fig:tw_triad_difference_roughness_wavenumber}
\end{figure}

Four travelling crossflow eigenmodes can interact with six roughness modes with suitable wavenumbers (figure \ref{fig:tw_resonance_mechanism}($a$)). A parametric study is to be performed for $m=0.34207$ and $m=1.0$ at $Re=338$ and $Re=2000$. As with stationary vortices, the wavenumbers of the roughness participating in the interactions are presented first with the particular attention to their `distances' to the wavenumbers $(\beta_s^r, \alpha_s^r)$ of the right-branch neutral stationary eigenmode. Then the shape functions of the roughness modes are displayed in order to monitor their near-resonant and non-resonant nature. Finally, the results on the growth-rate correction coefficient and amplitude ratio are presented. For brevity, the results on the roughness modes are shown only for $m = 1.0$; those for other values of $m$ are similar. 

Figure \ref{fig:tw_triad_sum_roughness_wavenumber} shows the roughness wavenumbers in the triad sum for the four chosen frequencies of travelling vortices with $m=1.0$ at $Re=338$ and $2000$. Among four possible triad-sum interactions, the wavenumbers of roughnesses 2+3 and 1+4 are fairly close to $(\beta_{s}^r,\alpha_{s}^r)$ for a considerable range of $(-\alpha_\hi)$, whereas those of 1+3 and 2+4 differ significantly from $(\beta_{s}^r,\alpha_{s}^r)$ except at the maximum growth rate, where roughnesses 1+3 and  2+3 have the same wavenumbers, and so do roughnesses 2+4 and 1+4 because eigenmodes 1 and 2 coalesce. 
The result suggests that the roughness modes 2+3 and 1+4 are mainly near-resonant, whereas 1+3 and 2+4 are mainly non-resonant. The interactions 2+3 and 1+4 are therefore of primary importance. 
As the Reynolds number increases, the wavenumbers of roughness 2+3 become closer to $(\beta_{s}^r,\alpha_{s}^r)$. 
As was mentioned earlier, when the frequency of travelling vortices tends to zero roughnesses 2+3 and 1+4 both become the roughness 1+2 for the stationary vortices governed by the triad sum, whereas roughnesses 1+3 and 2+4 become 1+1 and 2+2 governed by Bragg scattering. The near-resonant or non-resonant nature of the roughnesses in the triad sum for travelling vortices is consistent with that for stationary vortices.

Figure \ref{fig:tw_triad_difference_roughness_wavenumber} shows the roughness wavenumbers $(\beta_w, \alpha_w)$ in the triad difference for the four chosen frequencies, and clearly $(\beta_w, \alpha_w)$ differ significantly from $(\beta_{s}^r,\alpha_{s}^r)$ except near $-\alpha_\hi=0$. 
With the frequency increased, the roughness wavenumbers differ significantly from $(\beta_{s}^r,\alpha_{s}^r)$ even for $-\alpha_\hi\approx0$. The roughness modes 1-2 and 3-4 in the triad-difference interactions of travelling vortices are all non-resonant except for small $\omega$ and $-\alpha_\hi\approx0$. It is expected that the triad-difference interactions are in general less important than the triad-sum interactions. 

\begin{table}
  \begin{center}
\def~{\hphantom{0}}
  \begin{tabular}{lcccc}
              &$\omega$ & case 1    & case 2    & case 3         \\[3pt]
   $Re=338$   & 0.035   & 0.0076963 & 0.0038529 &  0.0000000          \\
   $Re=2000$  & 0.029   & 0.0094996 & 0.0047539 &  0.0000000   
  \end{tabular} 
  \caption{Growth rates for the chosen cases of travelling vortices at $m=1.0$.}
  \label{tab:tw_growth_rate}
  \end{center}
\end{table}

\begin{figure}
% \psfrag{y}{$y$}
% \psfrag{A}{$(a)$}
% \psfrag{B}{$(b)$}
% \psfrag{u}[][]{$|\hat{u}_w|$}
% \psfrag{v}[][]{$|\hat{v}_w|$}
% \psfrag{w}[][]{$|\hat{w}_w|$}
% \psfrag{l}{$y_c$}
\centerline{\includegraphics[scale=1]{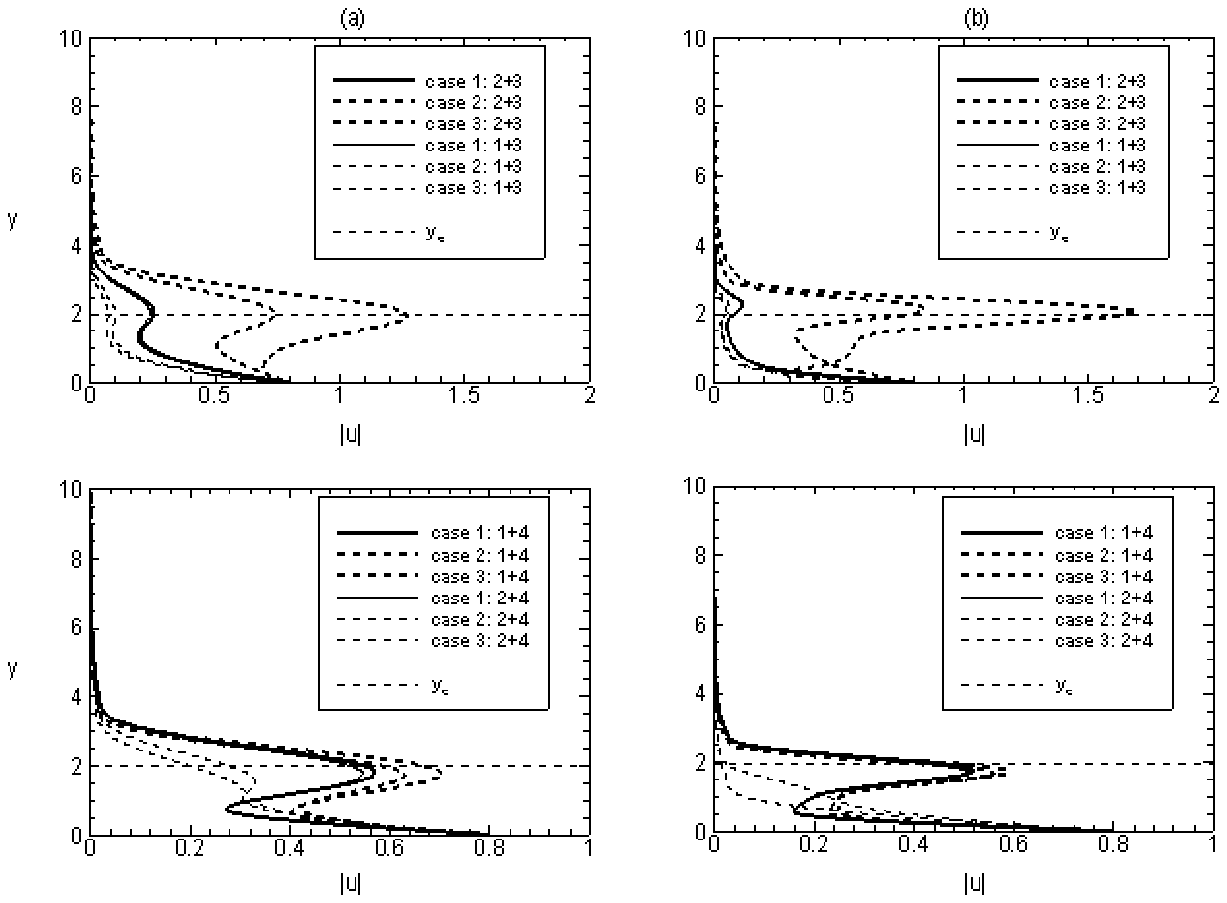}}
\caption{Distribution of the roughness modes in the triad-sum interactions with travelling vortices for $m=1.0$: $(a) Re=338$ and $\omega=0.035$; $(b) Re=2000$ and $\omega=0.029$.}
\label{fig:tw_triad_sum_shape_of_roughness}
\end{figure}

\begin{figure}
% \psfrag{y}{$y$}
% \psfrag{A}{$(a)$}
% \psfrag{B}{$(b)$}
% \psfrag{u}[][]{$|\hat{u}_w|$}
% \psfrag{v}[][]{$|\hat{v}_w|$}
% \psfrag{w}[][]{$|\hat{w}_w|$}
% \psfrag{l}{$y_c$}
\centerline{\includegraphics[scale=1]{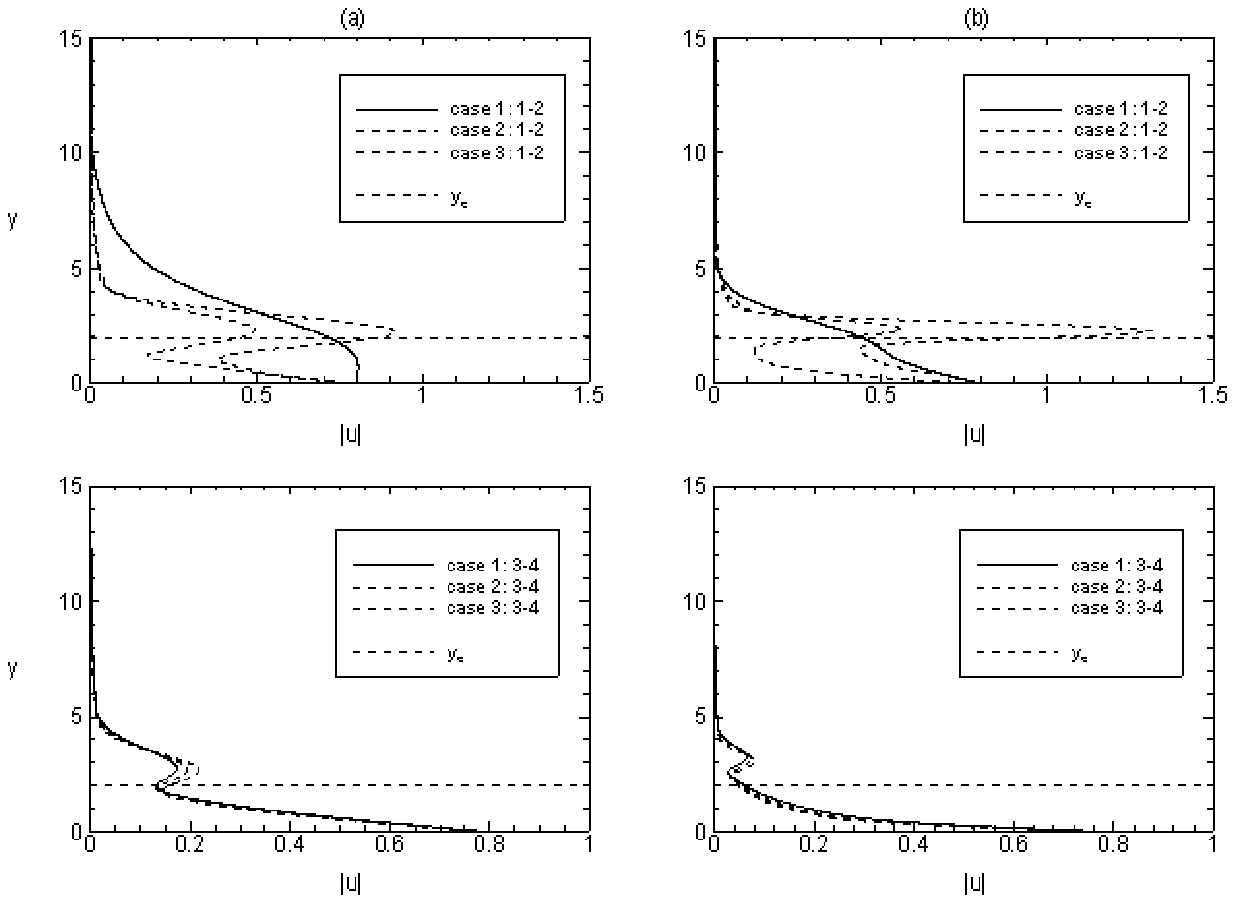}}
\caption{Distribution of the roughness modes in the triad-difference interactions with travelling vortices for $m=1.0$: $(a) Re=338$ and $\omega=0.035$; $(b) Re=2000$ and $\omega=0.029$.}
\label{fig:tw_triad_difference_shape_of_roughness}
\end{figure}

\begin{figure}
% \psfrag{x}[][]{$-\alpha_\hi$}
% \psfrag{y}{$\lambda_\hr$}
% \psfrag{A}[][]{$(a)$}
% \psfrag{B}[][]{$(b)$}
\centerline{\includegraphics[scale=1]{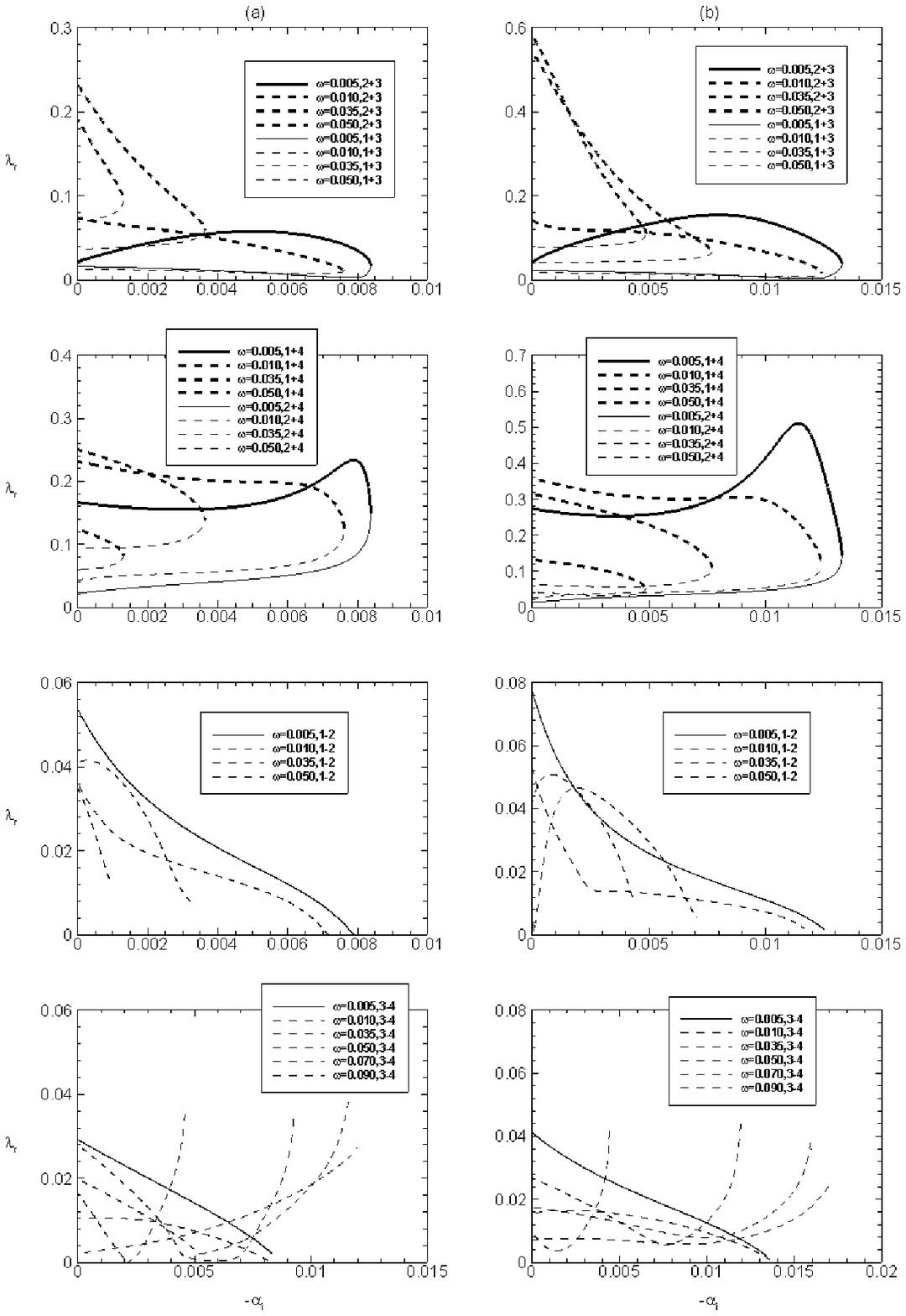}}
\caption{Variation of $\lambda_\hr$ versus $-\alpha_\hi$ for travelling vortices of different frequency at $Re=338$: $(a)$ $m=0.34207$; $(b)$ $m=1.0$.}
\label{fig:tw_re=338_lambda}
\end{figure}

\begin{figure}
% \psfrag{x}[][]{$-\alpha_\hi$}
% \psfrag{y}{$\lambda_\hr$}
% \psfrag{A}[][]{$(a)$}
% \psfrag{B}[][]{$(b)$}
\centerline{\includegraphics[scale=1]{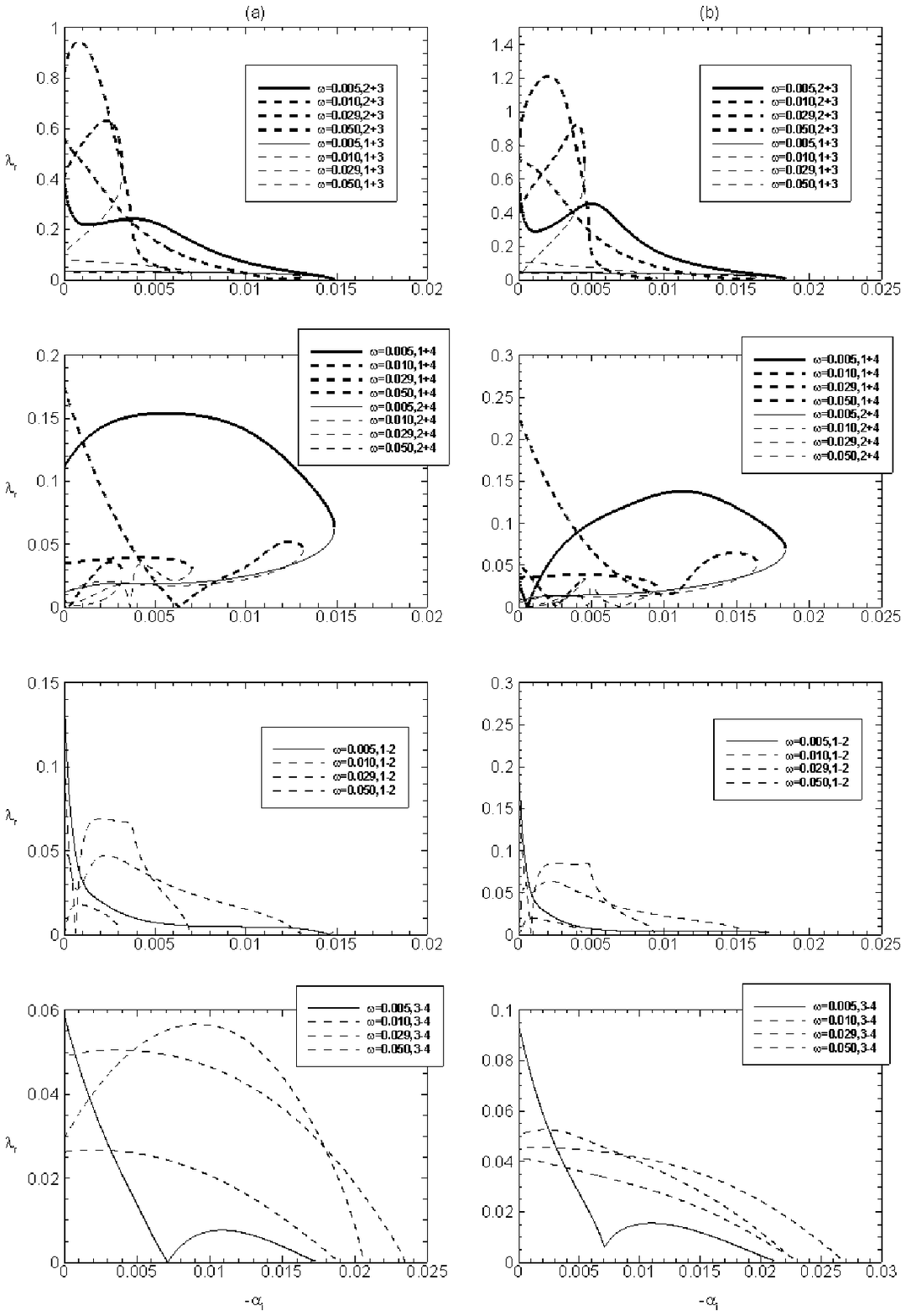}}
\caption{Variation of $\lambda_\hr$ versus $-\alpha_\hi$ for travelling vortices of different frequency at $Re=2000$: $(a)$ $m=0.34207$; $(b)$ $m=1.0$.}
\label{fig:tw_re=2000_lambda}
\end{figure}
\begin{figure}
% \psfrag{x}[][]{$-\alpha_\hi$}
% \psfrag{y}{$\lambda_r$}
% \psfrag{A}[][]{$(a)$}
% \psfrag{B}[][]{$(b)$}
% \psfrag{s}{$(1)$}
% \psfrag{n}{$(2)$}
\centerline{\includegraphics[scale=0.9]{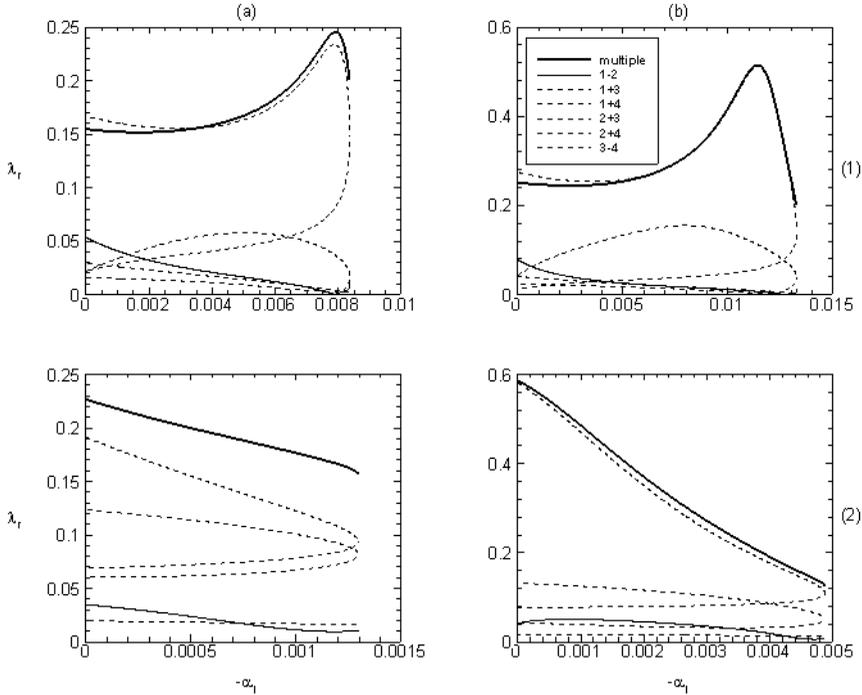}}
\caption{Comparisons of $\lambda_\hr$ for travelling vortices interacting with multiple-wavenumber and single-wavenumber roughness at $Re=338$: $(a)$ $m=0.34207$; $(b)$ $m=1.0$; $(1)$ $\omega=0.005$; $(2)$ $\omega=0.05$.}
\label{fig:tw_re=338_multiple_single_lambda}
\end{figure}

\begin{figure}
% \psfrag{X}[][]{$-\alpha_\hi$}
% \psfrag{x}[][]{$-\alpha_\hi$}
% \psfrag{y}{$\lambda_r$}
% \psfrag{A}[][]{$(a)$}
% \psfrag{B}[][]{$(b)$}
% \psfrag{P}{$(1)$}
% \psfrag{Q}{$(2)$}
% \psfrag{T}{$(3)$}
\centerline{\includegraphics[scale=0.9]{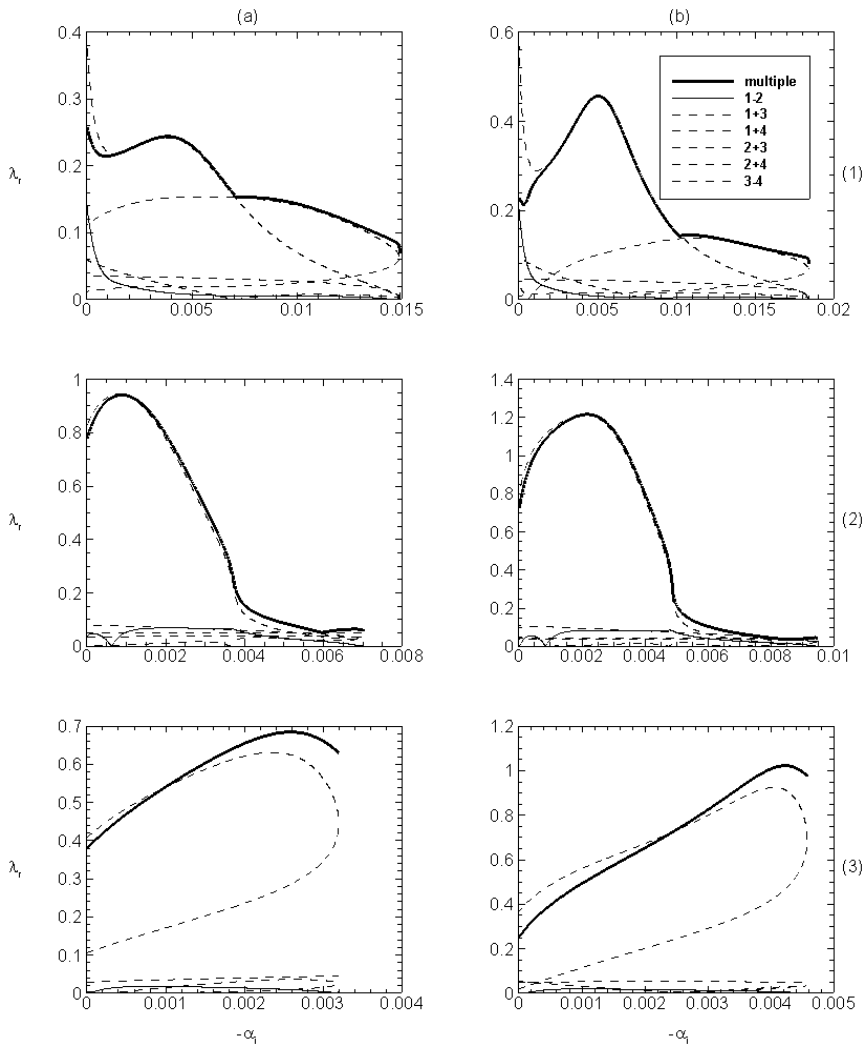}}
\caption{Comparisons of $\lambda_\hr$ for travelling vortices interacting with multiple-wavenumber and single-wavenumber roughness at $Re=2000$: $(a)$ $m=0.34207$; $(b)$ $m=1.0$; $(1)$ $\omega=0.005$; $(2)$ $\omega=0.029$; $(3)$ $\omega=0.05$.}
\label{fig:tw_re=2000_multiple_single_lambda}
\end{figure}
\begin{figure}
% \psfrag{x}[][]{$-\alpha_\hi$}
% \psfrag{y}{$\rho$}
% \psfrag{A}[][]{$(a)$}
% \psfrag{B}[][]{$(b)$}
\centerline{\includegraphics[scale=1]{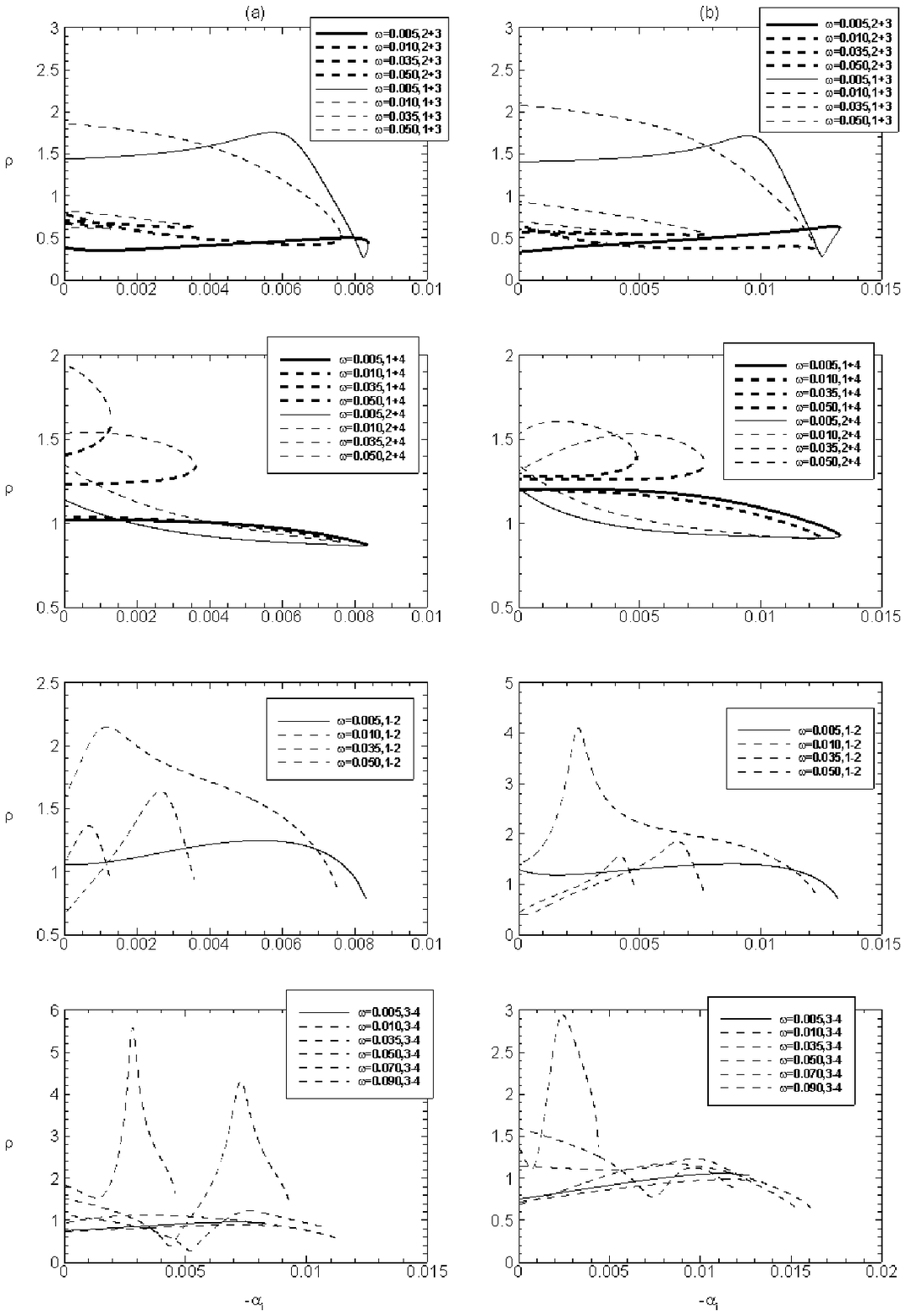}}
\caption{Variation of $\rho$ versus $-\alpha_\hi$ for travelling vortices of different frequency at $Re=338$: $(a)$ $m=0.34207$; $(b)$ $m=1.0$.}
\label{fig:tw_re=338_amplitude_ratio}
\end{figure}

\begin{figure}
% \psfrag{x}[][]{$-\alpha_\hi$}
% \psfrag{y}{$\rho$}
% \psfrag{A}[][]{$(a)$}
% \psfrag{B}[][]{$(b)$}
\centerline{\includegraphics[scale=1]{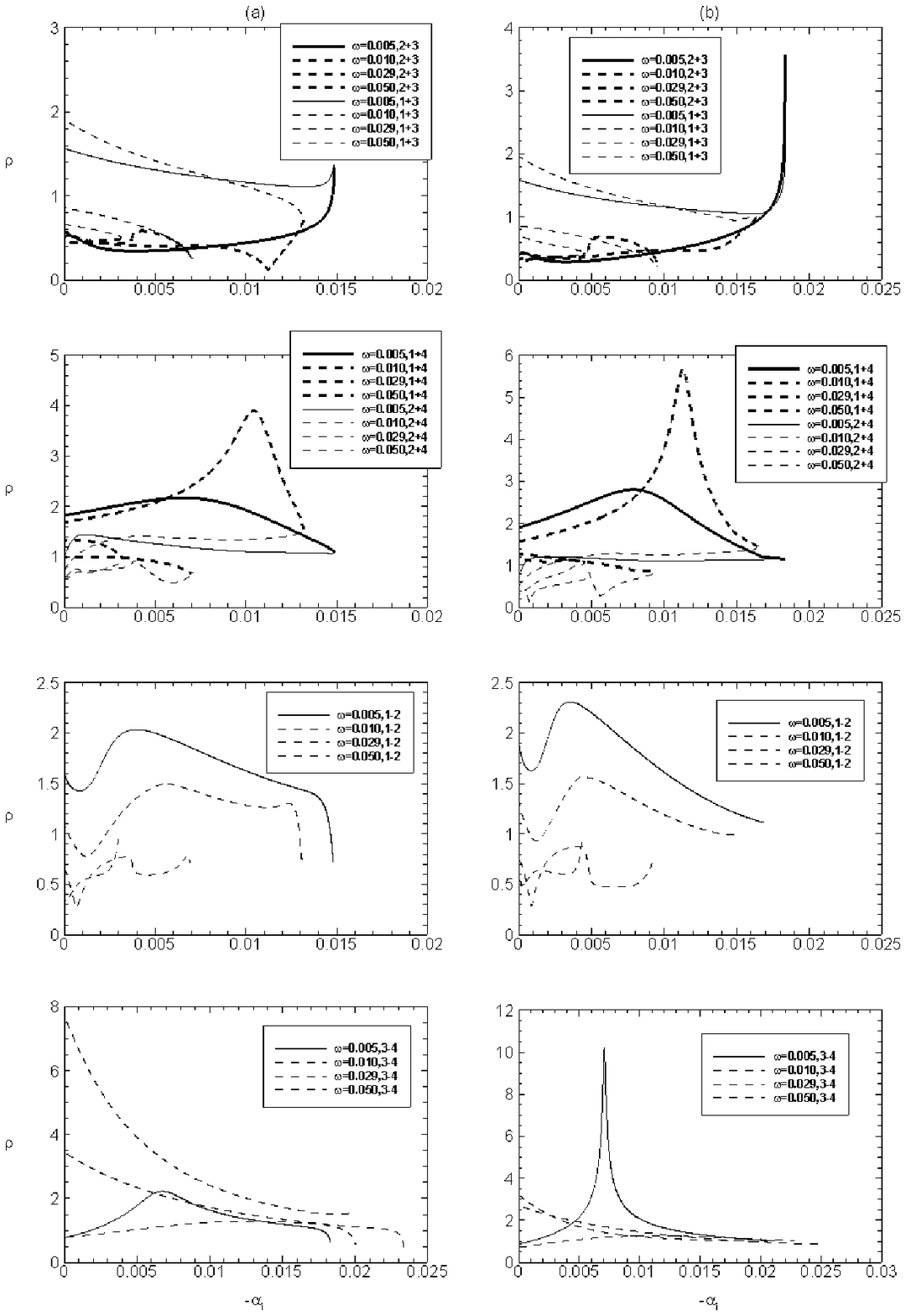}}
\caption{Variation of $\rho$ versus $-\alpha_\hi$ for travelling vortices of different frequency at $Re=2000$: $(a)$ $m=0.34207$; $(b)$ $m=1.0$.}
\label{fig:tw_re=2000_amplitude_ratio}
\end{figure}
\begin{figure}
% \psfrag{x}[][]{$-\alpha_\hi$}
% \psfrag{y}[][]{$\rho$}
% \psfrag{A}[][]{$(a)$}
% \psfrag{B}[][]{$(b)$}
% \psfrag{P}{$(1)$}
% \psfrag{Q}{$(2)$}
% \psfrag{M}{$|\frac{A_1}{A_4}|$}
% \psfrag{N}{$|\frac{A_2}{A_3}|$}
\centerline{\includegraphics[scale=0.85]{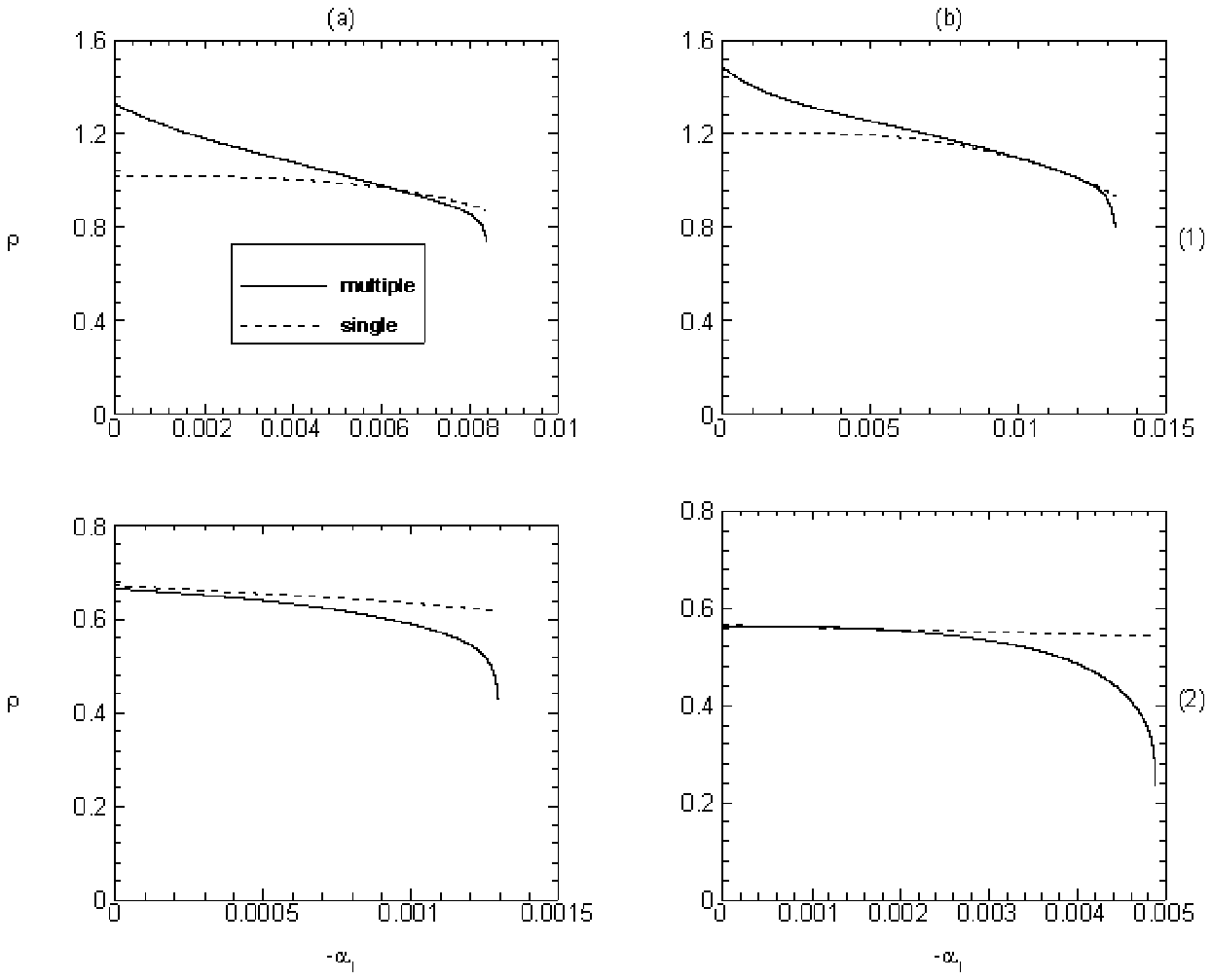}}
\caption{Comparison of the amplitude ratio in the multiple-wavenumber case with that in the dominant single-wavenumber case at $Re=338$: $(a)$ $m=0.34207$; $(b)$ $m=1.0$; $(1)$ $\omega=0.005$; $(2)$ $\omega=0.05$.}
\label{fig:tw_re=338_multiple_dominant_amplitude_ratio}
\end{figure}

\begin{figure}
% \psfrag{x}[][]{$-\alpha_\hi$}
% \psfrag{X}[][]{$-\alpha_\hi$}
% \psfrag{y}[][]{$\rho$}
% \psfrag{A}[][]{$(a)$}
% \psfrag{B}[][]{$(b)$}
% \psfrag{P}{$(1)$}
% \psfrag{Q}{$(2)$}
% \psfrag{M}{$|\frac{A_1}{A_4}|$}
% \psfrag{N}{$|\frac{A_2}{A_3}|$}
\centerline{\includegraphics[scale=0.85]{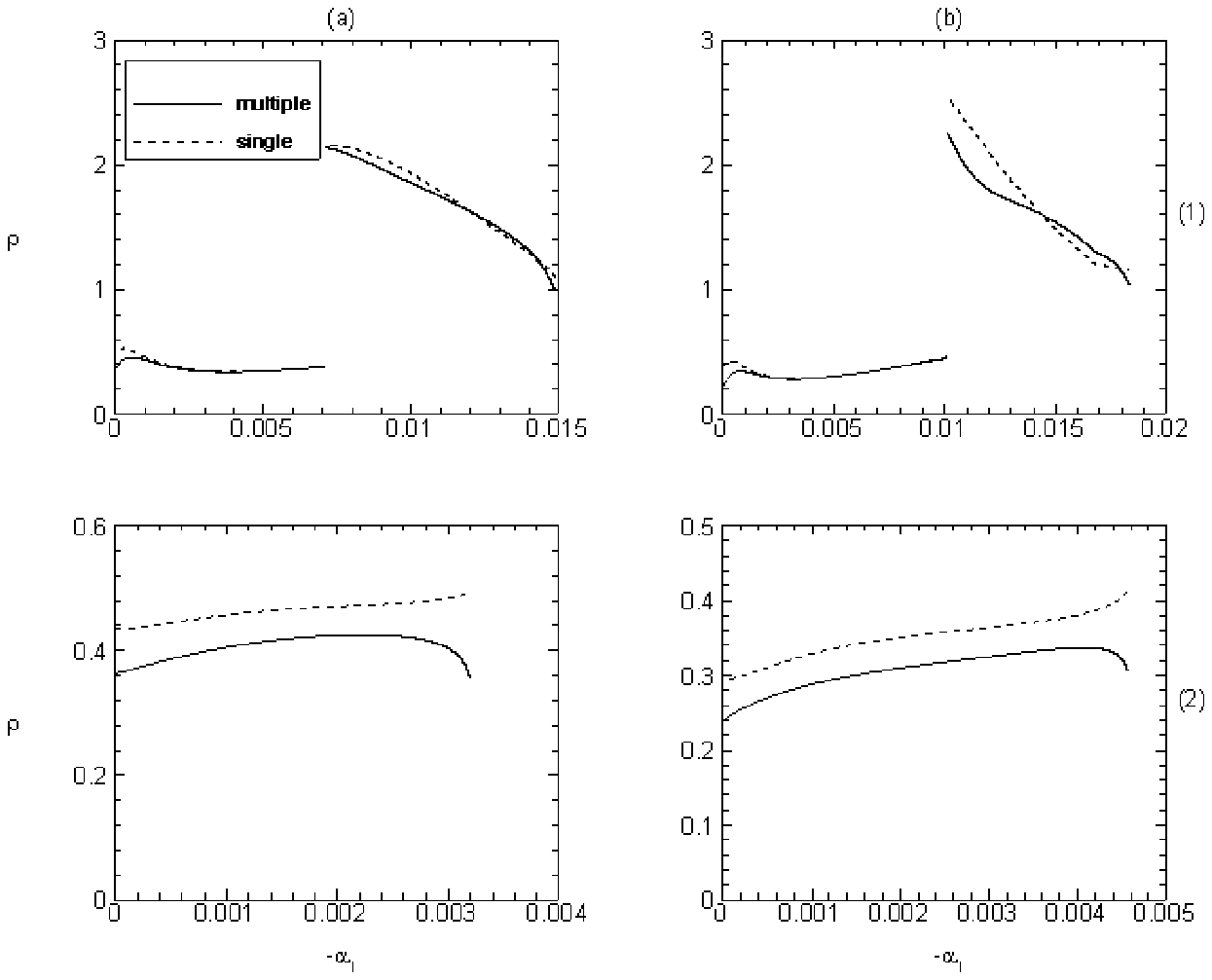}}
\caption{Comparison of the amplitude ratio in the  multiple-wavenumber case with that in the dominant single-wavenumber case at $Re=2000$: $(a)$ $m=0.34207$; $(b)$ $m=1.0$; $(1)$ $\omega=0.005$; $(2)$ $\omega=0.05$.}
\label{fig:tw_re=2000_multiple_dominant_amplitude_ratio}
\end{figure}

The shape functions of the roughness modes in the different interactions are now examined. The chosen frequencies for $Re=338$ and $Re=2000$ are 0.035 and 0.029 respectively, which are approximately the frequencies of the corresponding most unstable travelling vortices. For each $Re$ and $\omega$, three cases with different growth rates are chosen; see table \ref{tab:tw_growth_rate}. Among these, the growth rates of eigenmodes 1 and 2 in case 1 are close to the largest, and the eigenmodes in case 3 are neutral. Figure \ref{fig:tw_triad_sum_shape_of_roughness} shows the roughness modes in the triad sum at $Re=338$ and $2000$. The perturbation induced by roughness 2+3 is appreciably stronger than that induced by 1+3. This is a consequence of the fact that the wavenumbers of roughness 2+3 are closer to $(\beta_s^r, \alpha_s^r)$ than those of 1+3 (figure \ref{fig:tw_triad_sum_roughness_wavenumber}). The same trend can also be found for roughnesses 1+4 and 2+4. There exists a local peak near the critical level $y_c$. As the growth rate decreases from case 1 to case 3, the peak value rises for roughness 2+3 and falls for 1+3; the peak value for roughness 1+4 rises first and then falls, whereas that for roughness 2+4 falls. These behaviours are consistent with the closeness of the roughness wavenumbers to $(\beta_s^r, \alpha_s^r)$ in the triad sum as shown in figure \ref{fig:tw_triad_sum_roughness_wavenumber}.
At $Re=2000$, the near-resonant roughness modes display high-Reynolds-number features: the two-layer structure and the larger amplifications in the critical layer.

Figure \ref{fig:tw_triad_difference_shape_of_roughness} shows the signature of the roughness modes in the triad-difference interactions with travelling vortices at $Re=338$ and $2000$. According to the sizes of the roughness modes in the critical layer, only 1-2 of case 3 is near-resonant, and its peak value rises with the Reynolds number. As the growth rate increases from case 3 to case 2, the perturbation induced by roughness 1-2 decreases. The perturbations induced by roughness 3-4 for all three cases are all non-resonant. These results are also consistent with the `distances' of the roughness wavenumbers to $(\beta_s^r, \alpha_s^r)$ in the triad difference as indicated by figure \ref{fig:tw_triad_difference_roughness_wavenumber}.

Figure \ref{fig:tw_re=338_lambda} shows the growth-rate correction coefficient for travelling vortices in the triad sum and difference interactions at $Re=338$. For a fixed frequency of eigenmodes, the correction coefficients of roughnesses 2+3 and 1+3 are equal at the maximum growth rate because eigenmodes 1 and 2 coalesce there. The same remarks apply to roughnesses 1+4 and 2+4. The correction coefficient due to roughness 2+3 is larger than that due to 1+3. This can be attributed to the fact that the perturbations induced by the former are larger than those induced by the latter as shown in figure \ref{fig:tw_triad_sum_shape_of_roughness}($a$). For the same reason, the correction coefficient due to 1+4 is larger than that due to 2+4. 
For roughness 2+3 at high frequencies, as the growth rate decreases the correction coefficient increases.
For roughness 1+4 at $\omega=0.005$, there exists a prominent peak located close to the maximum growth rate for each acceleration parameter. The corresponding roughness wavenumbers are almost equal to $(\beta_s^r, \alpha_s^r)$ as shown in figure \ref{fig:tw_triad_sum_roughness_wavenumber}($a$), which means that the roughness mode is almost in exact resonance with the right-branch neutral stationary eigenmode. An exact resonance may take place in the non-stationary case under suitable conditions, as it does in the stationary case. 
The correction coefficients caused by 2+3 and 1+4 interactions in the triad sum are of  $\textit{O}(10^{-1})$, whereas those by 1+3 and 2+4 interactions are much smaller.  The correction coefficients due to the triad difference are of $\textit{O}(10^{-2})$. The reason for this significant difference is that the perturbations induced by roughnesses 2+3 and 1+4 are mainly near-resonant, while those induced by roughnesses 1+3, 2+4, 1-2 and 3-4 are non-resonant, as was demonstrated earlier. Overall, as $m$ increases, so does the correction coefficient. The correction coefficient at $Re=2000$ is shown in figure \ref{fig:tw_re=2000_lambda}. The correction coefficient due to roughness 2+3 for low growth rates and intermediate frequencies can be extraordinarily large, owing again to the fact that the corresponding roughness wavenumbers are close to $(\beta_s^r, \alpha_s^r)$ (figure \ref{fig:tw_triad_sum_roughness_wavenumber}($b$)) and the roughness modes are amplified hugely in the critical layer (figure \ref{fig:tw_triad_sum_shape_of_roughness}($b$)). This result suggests that the leading-order growth rate cannot capture the spatial instability of travelling vortices in the presence of roughness, and the effect of roughness must be considered. 

The correction coefficients for travelling vortices interacting with multiple-wavenumber and single-wavenumber roughnesses are compared in figures \ref{fig:tw_re=338_multiple_single_lambda} and \ref{fig:tw_re=2000_multiple_single_lambda} for $Re=338$ and $Re=2000$ respectively. The multiple-wavenumber roughness is comprised of the six Fourier components all having equal amplitude. Of the six resonant interactions, there exists a dominant one, which makes the largest contribution for each set of parameters. Broadly speaking, the dominant roughness is 2+3 or 1+4, each driving a perturbation in a near-resonant manner. Specifically,  at $Re=338$ the dominant single-wavenumber roughnesses are 1+4 and 2+3 for low and high frequencies respectively. At a high Reynolds number ($Re=2000$), for low frequency ($\omega=0.005$) there exists an intersection between the correction coefficients due to roughnesses 1+4 and 2+3 as shown in figure \ref{fig:tw_re=2000_multiple_single_lambda}(1); roughnesses 2+3 and 1+4 play a dominant role at low and high growth rates respectively.
For an intermediate frequency ($\omega=0.029$) (figure \ref{fig:tw_re=2000_multiple_single_lambda}(2)), roughness 2+3 dominates at almost all growth rates. For a high frequency ($\omega=0.05$) (figure \ref{fig:tw_re=2000_multiple_single_lambda}(3)), the dominant single-wavenumber roughness is always 2+3. The correction coefficient $\lambda_\hr$ due to multiple-wavenumber roughness is of a similar size to that due to the dominant single-wavenumber roughness, and typically $\lambda_\hr=\textit{O}(10^{-1})$, which implies that micron-sized roughness with multiple wavenumbers can create an $\textit{O}(1)$ correction to the growth rates of travelling vortices as we will show later. For larger acceleration parameter $m$, the correction coefficient due to multiple-wavenumber roughness increases. The correction coefficient at low and high Reynolds numbers exhibits different features: for low-frequency vortices large corrections at $Re=338$ occur at high growth rates, while those at $Re=2000$ occur at low growth rates; for high-frequency vortices the correction coefficient at $Re=338$ decreases with the growth rate, while that at $Re=2000$ generally increases with the growth rate.

The other effect of distributed roughness on crossflow eigenmodes through the generalized resonance mechanisms is that it renders the eigenmodes involved in the resonance to obey a fixed amplitude ratio. The variation of the amplitude ratio $\rho$ with the growth rate for travelling vortices of different frequency is shown in figures \ref{fig:tw_re=338_amplitude_ratio} and \ref{fig:tw_re=2000_amplitude_ratio} for $Re=338$ and $Re=2000$ respectively. Note that for roughness $i+j$ or $i-j$, $\rho$ is taken as the ratio of the amplitude of eigenmode $i$ to that of eigenmode $j$. At $Re=338$ for dominant interactions $\rho$ takes moderate values in the range of $0.5\leq\rho\leq2$, but for certain non-dominant interactions, e.g. 3-4, $\rho$ can be quite large. At $Re=2000$, $\rho$ can be quite large as well for dominant interactions of relatively low frequency vortices. 

As noted earlier, among all possible interactions, one is dominant in the sense that it more-or-less dictates the growth-rate correction. One may ask whether the same behaviour extends to the amplitude ratio. This question is answered in figures \ref{fig:tw_re=338_multiple_dominant_amplitude_ratio} and \ref{fig:tw_re=2000_multiple_dominant_amplitude_ratio} for $Re=338$ and $Re=2000$ respectively. At $Re=338$, $\rho$ in the multiple-wavenumber case follows that in the single-wavenumber case for low frequencies at high growth rates and for high frequencies at low growth rates. At $Re=2000$, there is a close agreement. The discontinuity of $\rho$ with the growth rate for $\omega=0.005$ (figure \ref{fig:tw_re=2000_multiple_dominant_amplitude_ratio}($1$)) occurs where the dominant interaction changes as shown in figure \ref{fig:tw_re=2000_multiple_single_lambda}(1).  Therefore, we can conclude that in the multiple-wavenumber case, generally there exists a dominant interaction in the sense that the corresponding growth-rate correction coefficient and amplitude ratio are both marginally affected by other roughness components.

Our numerical results show that the growth-rate correction coefficient $\lambda_\hr$ caused by the resonant interactions between the crossflow eigenmodes and roughness modes is about $0.3$ in the majority of the parameter space. The correction to the leading-order growth rate is $\lambda_\hr h$ with the relative roughness height being $h=h^*/\delta^*_0$. The typical value of $h$ for a micron-sized roughness is now estimated by referring to the experimental data of \citet{radeztsky1999effect}. In their experiment, a $12\mu m$ isolated roughness element was placed at $x^*/c^*=0.023$, and the non-dimensionalized height is defined as $h^*/\delta^*_{0.99}=0.014$ with $\delta^*_{0.99}$ representing the transverse position where $U=0.99$. Since $x^*/c^*=0.023$ is fairly close to the leading edge, the local boundary layer can be approximated by the FSC solution with $m=1.0$.
It follows that $h^*/\delta^*_0$ can be decided by $h^*/\delta^*_0=(h^*/\delta^*_{0.99})y_{0.99}$, where $y_{0.99}=\delta^*_{0.99}/\delta^*_0$ is given by $U(y_{0.99})=0.99$. For the FSC boundary layer with $m=1.0$, our computation gives $y_{0.99}=3.6725$, and it follows that ${h^*}/{\delta^*_0}=0.0514$ and $\delta^*_0=0.2334mm$. The chord Reynolds number $Re_c=U_{\infty,s}^*c^*/\nu^*=2.6\times 10^6$, where $U_{\infty,s}^*$ is the streamwise free-stream velocity, and the airfoil chord length $c^*=1.83m$ \citep{carrillo1997distributed}. The sweep angle in the experiment is set to $45^\circ$. The Reynolds number based on $\delta^*_0$ at $x^*/c^*=0.023$ is therefore estimated as $Re=(Re_c\delta^*_0)/(\sqrt{2}c^*)=234.5$, fairly close to the low Reynolds number (338) in our computation. The Reynolds number would be equal if the roughness is present somewhat downstream where $\delta^*_0=0.2334\times338/234.5\approx0.34mm$. Hence, the relative height ($h=h^*/\delta^*_0$) of the micron-sized roughness (typically $h^*\approx10\mu m$ for unpolished surfaces) at the lower Reynolds number ($Re=338$) in the experimental setting is around $0.03$. The correction to the growth rate of both stationary and travelling vortices is roughly $0.01$, which is comparable with the leading-order growth rates (\textit{O}($10^{-2}$)).

\section{Conclusions and future work}\label{sec:conclusions}
The present paper sought to provide a theoretical explanation for the extreme sensitivity of transition in three-dimensional boundary layers to micron-sized distributed roughness. Unlike much of the existing work which focuses on the receptivity process or local instability of the modified base flow, we approached this problem by studying the effect of distributed surface roughness on crossflow instability through resonant interactions of crossflow eigenmodes with the roughness-induced perturbations. With the latter being represented as a superposition of different Fourier components, we identified several resonances which can take place between one or more growing eigenmodes and suitable roughness components to affect the growth rate of crossflow vortices. These include Bragg scattering involving a stationary eigenmode and a roughness mode, and resonant-triad interactions between two eigenmodes and a roughness component. The latter may take two forms: triad sum and triad difference, which refer to the cases where the sum/difference of the wavenumbers of the eigenmodes is equal to those of the roughness, respectively. The present Bragg scattering and resonant-triad interactions are generalized in that the eigenmodes have $\textit{O}(1)$ growth rates. For the triadic interaction, all that is required is that the two eigenmodes must have equal growth rate. Owing to the resonant nature of the interactions, the correction to the growth rate is proportional to the roughness height and the scaling factor is defined as the growth-rate correction coefficient. By a multi-scale method, we derive the amplitude equations of the crossflow eigenmodes interacting with distributed surface roughness through the generalized resonance mechanisms. After solving the amplitude equations, we obtained the growth-rate correction coefficient $\lambda_\hr$ and amplitude ratio $\rho$ of the participating eigenmodes in the resonances. 

The fact that the amplitude ratio $\rho$ is fixed by the roughness means that the eigenmodes involved in the resonance are fully coupled at the leading order; this is in stark contrast to the smooth-wall case where eigenmodes in the linear regime are independent of each other with their amplitude ratio being arbitrary. An important implication is that the spectral composition of crossflow vortices in a 3D boundary layer is affected by distributed surface roughness. 

The size of $\lambda_\hr$ depends on the `distance' between the roughness wavenumbers $(\beta_w, \alpha_w)$ and the wavenumbers $(\beta_s^r, \alpha_s^r)$ of the right-branch neutral stationary eigenmode, and the roughness modes can be categorized as being near-resonant if the `distance' is small and non-resonant otherwise. The numerical results for the FSC boundary layer indicate that, for a fairly wide range of parameters, the roughness wavenumbers satisfying one of the resonance conditions turn out to be close to $(\beta_s^r, \alpha_s^r)$, leading to a substantial correction to the growth rate. The corresponding length scales might be what \citet{radeztsky1999effect} referred to as `dangerous scales' that distributed roughnesses seemed to have. The numerical evaluations show that at moderate Reynolds numbers the nearly most unstable stationary vortices are most sensitive to roughness through Bragg scattering. The growth-rate correction coefficient due to the triad sum for stationary vortices is generally larger than that due to the triad difference because the participating roughness mode is near-resonant in the former but non-resonant in the latter. For travelling vortices, two triad-sum interactions were found to be of primary importance in the respective ranges of the parameters. 
Multiple-wavenumber roughness, which is more realistic than its single-wavenumber counterpart, produces correction coefficients similar to what the dominant single-wavenumber roughness does. In the majority of the parameter space, the growth-rate correction coefficient is about $0.3$, which means that roughness with $10\mu m$ height induces a more-or-less $\textit{O}(1)$ correction to the leading-order growth rates of both stationary and travelling vortices at typical Reynolds numbers where the boundary-layer displacement thickness is about $1mm$. Therefore, the generalized resonant interactions appear to offer a possible explanation for the experimentally observed high sensitivity of transition in three-dimensional boundary layers to micron-sized distributed roughness.

The mechanisms identified in the present paper are promising enough that further work on them is warranted. It is of interest to investigate the resonant interactions in realistic three-dimensional boundary layers over swept wings. We believe that the same mechanisms operate and remain significant there, given the fact that the existence of the resonant interactions relies upon only the generic feature of the dispersion relation of crossflow instability, and their effects are significant for different values of the acceleration parameter in the FSC family, as was demonstrated in the present paper. We have assumed that all Fourier components in the surface roughness have constant heights, which allows the system of the amplitude equations to be reduced to an eigenvalue problem, but the analysis can be extended to roughnesses whose heights are modulated in space, e.g. are functions of $\bar{x}=hx$, for which the amplitude equations would have non-constant coefficients, and will have to be investigated as an initial-value problem. It was demonstrated that micron-sized roughness can potentially alter the growth rate by an $\textit{O}(1)$ amount, but one might question whether the perturbation method used would remain appropriate when the correction is not small anymore. Direct numerical simulations may thus be necessary. In the present paper, the crossflow vortices are assumed linear, whereas effects of distributed roughness on nonlinear vortices await to be studied. Finally, we hope that the present theoretical results could prompt further experimental efforts to quantify the effects of distributed roughness.

\section*{Acknowledgement}
The computer code used for the present work was developed from the one generously supplied by late Prof. J. S. Luo of Tianjin University.  XW would like to thank Prof. W. S. Saric and Dr. M. Choudhari for helpful discussions on the role of surface roughness in three-dimensional boundary-layer transition.

\bibliographystyle{jfm}
% \bibliography{he_wu}

\end{document}